\documentclass{article}
\usepackage{graphicx} % Required for inserting images

\usepackage[super,sort&compress,comma]{natbib} 
\usepackage[version=3]{mhchem}
\usepackage[left=2.5cm, right=2.5cm, top=1.785cm, bottom=2.0cm]{geometry}
\usepackage{balance}
\usepackage{mathptmx}
\usepackage{sectsty}
\usepackage{lastpage}
\usepackage[format=plain,justification=justified,singlelinecheck=false,font={stretch=1.125,small,sf},labelfont=bf,labelsep=space]{caption}
\usepackage{float}
\usepackage{fancyhdr}
\usepackage{fnpos}
\usepackage[english]{babel}
\addto{\captionsenglish}{%
}
\usepackage{array}
\usepackage{droidsans}
\usepackage{charter}
\usepackage[T1]{fontenc}
\usepackage[usenames,dvipsnames]{xcolor}
\usepackage{setspace}
\usepackage[compact]{titlesec}
\usepackage{hyperref}
\usepackage{siunitx}

\usepackage{epstopdf}%This line makes .eps figures into .pdf - please comment out if not required.

\title{Quantifying the liquid-liquid transition in cold water/glycerol mixtures by ih-RIDME}
\author{Sergei Kuzin$^{1*}$ and Maxim Yulikov$^{1*}$}
\date{September 2025}

\begin{document}

\maketitle
$^{1}$ Department of Chemistry and Applied Biosciences, Vladimir-Prelog-Weg 2, 8093, Zurich, Switzerland\\[1em]
*E-mail: sergei.kuzin@mpinat.mpg.de\footnote[2]{Current affiliation: Max Planck Institute for Multidisciplinary Sciences, Am Fassberg 11, 37077 G\"ottingen, Germany.}, myulikov@ethz.ch

\section{Abstract}

Water/glycerol mixtures are common for experiments with biomacromolecules at cryogenic temperatures due to their vitrification properties. Above the glass transition temperature, they undergo liquid-liquid phase separation. Using the novel EPR technique called intermolecular hyperfine Relaxation-Induced Dipolar Modulation Enhancement (ih-RIDME), we quantified the molar composition in frozen water/glycerol mixtures with one or the other component deuterated after the phase transition. Our experiments reveal nearly equal phase composition regardless of the proton/deuterium isotope balance. With the new ih-RIDME data, we can also revisit the already reported body of glass transition data for such mixtures and build a consistent picture for water/glycerol freezing and phase transitions. Our results also indicate that ih-RIDME has the potential for investigating the solvation shells of spin-labelled macromolecules.

\section{Introduction}

For deeply cooled or frozen solvent mixtures, knowing the local phase state, local composition, as well as possibly inhomogeneous local density distributions for each individual molecular type, besides the fundamental interest, can also be essential for performing advanced spectroscopic experiments. In particular, this applies to magnetic resonance: different types of solid-state (SS) nuclear magnetic resonance (NMR),\cite{Bauer2017LTSSNMR,Meier2022FastMASHepVB, Oschkinat2011CryoTResolution,McDermott2012SSNMRLinewidth} dynamic nuclear polarisation (DNP),\cite{OschkinatLesage2019DNPbioNMR,DePaepe2015NMRenhancedbyDNP} or pulse electron paramagnetic resonance (EPR).\cite{bordignon2015_glasses,emmanouilidis2021nmr} Similar considerations might also appear important beyond the magnetic resonance spectroscopy, in any other spectroscopy, crystallography or microscopy measurements at low temperatures, e.g. in the currently boosting field of cryo-electron microscopy.\cite{Garman2006,CryoEM01,CryoEM02} 

In magnetic resonance and beyond, solvents and solvent mixtures that can form glass upon shock freezing are of practical importance. The solvent mixture must form an amorphous bulk phase, without major solvent crystallisation. This prevents EPR- or NMR-active solutes from locally concentrating or even precipitating, so that the distribution of the dissolved molecules remains homogeneous. With this condition fulfilled, EPR and NMR can be used to study the molecular organisation of the glass around the probes.

The structure, free energy and composition of the solvation shells at ambient temperature have been addressed in the last few years by THz calorimetry and by water dynamics measurements with Overhauser DNP.\cite{SongiHan2024ODNP_SolventRestructuring,SongiHan2024WaterPolymerDiffusivity,Havenith2023LocalSolvationWaterGlycerol,Havenith2024HydrationFibrils,Havenith2024SolvationLLPS,Havenith2025TuningByCoSolutes} In particular, water dynamics in water/glycerol mixtures at ambient temperatures have been studied.\cite{SongiHan2024ODNP_SolventRestructuring} Also, for such mixtures at ambient temperatures, contributions of bulk water, wrap water (around hydrophobic moieties) and bound water were evaluated, and the Gibbs free enthalpy, the entropy and the enthalpy of mixing were determined.\cite{Havenith2023LocalSolvationWaterGlycerol} This set of data allows for a better verification of the MD computations quality. Also, the comparison of Overhauser DNP, THz calorimetry and MD enabled the discussion of coordination motifs (tetrahedral, icosahedral) in the water/glycerol mixtures.\cite{SongiHan2024ODNP_SolventRestructuring,Havenith2023LocalSolvationWaterGlycerol} However, is the molecular organisation of the glassy phase around large solutes like proteins, nucleic acids or polysaccharides upon freezing close to that at ambient temperature? Normally, a positive answer is assumed as it allows for the transfer of low-temperature data to the native conformational state of macromolecules. On the other hand, glass is a metastable state; therefore, it can change with time and affect the solute. It is attractive to address these questions by pulse EPR, as it offers a variety of powerful characterisation methods. Freezing samples for pulse EPR measurements changes the balance between entropic and enthalpic contributions to the chemical potential of each type of solute and solvent molecules. The corresponding changes in the solvent mixture composition might appear or might be, e.g. kinetically hindered. Following such changes would be useful for shedding light on the room temperature solvation properties.

Here, we worked with frozen water/glycerol mixtures, which are most commonly used for studying biomacromolecules at cryogenic temperatures. However, one of the standard storage conditions for protein and nucleic acid solutions is at $T=-80^{\circ}\,$C (193 K). This is close to the glass transition temperature, estimated at $158\pm5$~K by differential scanning calorimetry for pure water/glycerol mixture with 50\% (w/w) glycerol fraction\cite{angarita2021revisiting} and in the range 164-190 K from EPR data with spin probes and spin-labelled protein molecules at a ca. 50\% volume fraction of glycerol.\cite{bordignon2015_glasses} In general, considering any glass-forming solvent mixture, frozen samples stored above the glass transition temperature ($T_g$) exist in the form of a supercooled liquid. This is a state of matter between glass transition and melting temperatures, characterised by the absence of long-range order and high but finite viscosity. Under these conditions, the modification of the structure of the deeply cooled liquid may occur, some mechanical stress may relax, and a fraction of the solvent may partially crystallise or undergo other phase transitions. These processes are referred to as glass ageing.\cite{bohmer2024time,amir2012relaxations} It was also found that above the glass formation temperature, deeply cooled water/glycerol mixtures can undergo isocompositional liquid-liquid transition (LLT) into glycerol-rich (Liquid I) and water-rich (Liquid II) phases, without macroscopic phase separation.\cite{murata2012liquid} Such an LLT has been reported to be mainly driven by the local restructuring of water rather than glycerol.

In Ref.~\cite{bordignon2015_glasses}, the local rearrangement of hydrogen bonds formed between solvent molecules and the spin labels has been observed in water/glycerol around the glass transition temperature.\cite{bordignon2015_glasses} The recently introduced intermolecular hyperfine relaxation-induced dipolar modulation enhancement (ih-RIDME) technique\cite{kuzin2022diffusion,kuzin2024quantification,kuzin2025ihridme} is well suited to probe the composition and inhomogeneities of solvent molecules distributions around paramagnetic centres, with sensitivity region well beyond the first solvation shell, thus complementing the previously published hydrogen bonding data and revealing the overall composition of solvent in the spin probe's vicinity.

In the present paper, we compare the ih-RIDME data for protonated and deuterated water/glycerol mixtures with the data from more conventional pulse EPR techniques, such as Hahn echo decay and the matrix-peak electron spin echo envelope modulation (ESEEM). With these pulse EPR data, we can quantify the LLT and assign nitroxide spin probes to persist in the glycerol-rich phase. These experiments exemplify the capability of ih-RIDME for the quantitative determination of the distribution of solvent molecules in a local surrounding of spin probes, which has been earlier evaluated to cover up to 3 nm distances from the spin probe,\cite{kuzin2022diffusion,kuzin2025ihridme,kuzin2025ridme} depending on the solvent deuteration level. This also allows us to propose the ih-RIDME technique as a pulse EPR experiment for investigating the solvation of biopolymers.

\section{Theoretical background for ih-RIDME}

\subsection{Pulse sequence}
\begin{figure*}
    \centering
    \includegraphics[width=0.9\linewidth]{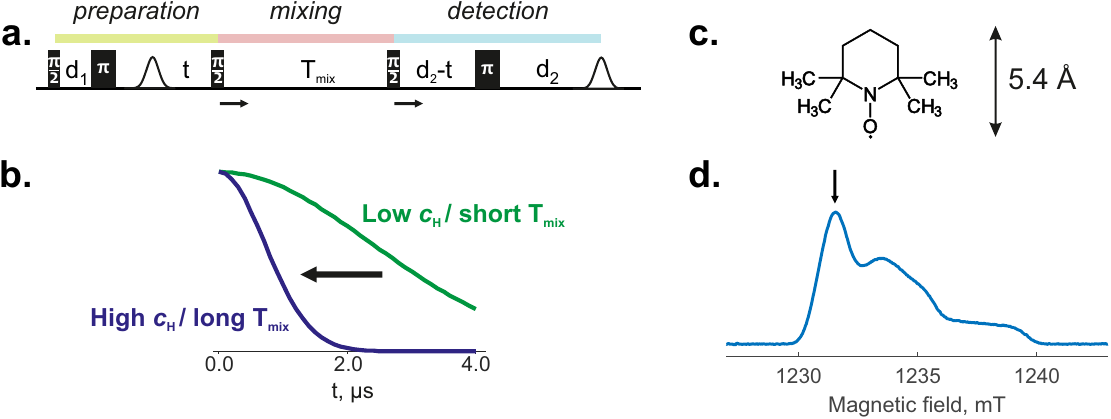}
    \caption{a. Five-pulse ih-RIDME pulse sequence with marked preparation, mixing and detection parts. Pulses are labelled by their flip angles. The mixing block is progressively shifted, causing echo decay. b. The decay of the ih-RIDME traces is steeper at higher proton concentrations or at longer mixing times. c. Chemical structure of nitroxide radical TEMPO used in this work. All protons in this molecule are within the frozen sphere and are not observable in ih-RIDME.\cite{kuzin2022diffusion} d. Echo-detected EPR spectrum of TEMPO in freshly frozen solution 2. The arrow indicates the positions of all EPR measurements.}
    \label{fig:01}
\end{figure*}

The pulse sequence for the five-pulse ih-RIDME experiment is shown in Figure~\ref{fig:01}a.\cite{kuzin2025ridme} In contrast to the conventional RIDME technique,\cite{milikisyants2009pulsed} which uses the same pulse sequence but relies on the spontaneous spin flips in a pair of coupled electron spins, the ih-RIDME experiment is designed to probe the interaction of a single electron spin with a `cloud' of protons around it. A favourable condition for the ih-RIDME is low concentration of the spin probe ($<100~\si{\micro M}$) to avoid strong electron spin-spin contributions.\cite{keller2019intermolecular} Where dilution of electron spins is challenging, the suppression of the electron-electron contributions can be aided by selecting lower measurement temperatures so that the typical longitudinal relaxation times of spin probes substantially exceed the mixing times necessary for the ih-RIDME. This step may involve a compromise on sensitivity.

The pulse sequence of the ih-RIDME experiment consists of preparation, mixing and detection blocks. The preparation block first creates a primary echo to avoid the dead-time problem.\cite{milikisyants2009pulsed} The electron coherence, refocused at the point of the echo, continues evolving for time $t$ under the hyperfine field exposed by close-by magnetic nuclei. The first $\pi/2$-pulse of the mixing block converts coherence into a polarisation grating along the static magnetic field. After the mixing time $T_{\textrm{mix}}$, the second $\pi/2$-pulse of the mixing block turns the polarisation grating back to the transversal plane. This coherence is then refocused by the second $\pi$-pulse and forms the detected echo. The amplitude of the echo gradually decreases with increasing time $t$ as well as with increasing mixing time $T_{\textrm{mix}}$ (Figure~\ref{fig:01}b). The following subsection introduces the mathematical model for this behaviour.

\subsection{Formalism of hyperfine spectral diffusion}
It is generally known that the simultaneous evolution of the electron-nuclear spin system under hyperfine and homonuclear dipolar coupling leads to the spin decoherence phenomenon in pulse EPR.\cite{deSousa2003, Witzel2006,jahn2022mechanism,jeschke2023nuclear,kuzin2024perturbation} The effect of the homonuclear coupling can be simplified as continuous flip-flop-like oscillations in the closest spin pairs. Thus, it mixes the hyperfine levels and makes the hyperfine energy non-stationary after the electron spin is excited. For the same reason, the polarisation grating is non-stationary during the mixing block. In the nuclear spin ensemble, hyperfine energy decorrelates, and this collective dynamics can be effectively described in the formalism of electron spin spectral diffusion, which comprises the core of the analytical model for ih-RIDME.

As shown in the analytical solution with two nuclei, the nutation frequency depends on both homonuclear coupling and the gradient of the hyperfine field within a nuclear pair.\cite{jeschke2023nuclear} In amorphous matter with many protons, internuclear vectors are uniformly distributed over a sphere and, to a good approximation, uncorrelated from electron-nuclear vector orientation and length. Consequently, the frequencies of such nuclear pair nutations are broadly distributed. Also, due to the weak correlation of electron-nuclear and nuclear-nuclear interaction in the unstructured spin bath,\cite{kuzin2024perturbation} the spectral diffusion kinetics allows a description by dissipative models, despite being formally a coherent process.

In the formalism of hyperfine spectral diffusion, homonuclear interaction is considered a perturbation to the hyperfine coupling. We introduce a state density function of multi-nuclear hyperfine levels, $\rho(\omega)$, also called the hyperfine spectrum. Each spin packet with hyperfine frequency $\omega$ obtains the phase $\omega t$ during transverse evolution. This is described by an ensemble function $\mu_t(\omega)=\rho(\omega)\exp(i\omega t)$ called magnetization spectrum. For a realistic number of about 100 nuclei in the electron spin's vicinity, we deal with a high number $2^{100}\sim 10^{30}$ of closely spaced nuclear states that form a quasi-continuous multi-spin hyperfine band. Thus, we can approximate $\rho(\omega)$ by a smooth function. This allows us to apply the formalism of continuous fluctuations for the hyperfine field during the mixing block and thereby describe its contribution to the ih-RIDME decay by a diffusion-like equation\cite{kuzin2022diffusion}
\begin{equation}\label{eq:master}
    \frac{\partial \mu_t(\omega, T)}{\partial T}=D\left(\rho(\omega)\frac{\partial^2 \mu_t(\omega, T)}{\partial \omega^2}-\rho''(\omega)\mu_t(\omega, T)\right)\,.
\end{equation}
Here, $T$ is time after the beginning of the mixing block, and $D$ is the spectral diffusion coefficient. After solving this equation of motion for magnetisation spectrum during the mixing block with initial condition $\mu_t(\omega, 0):=\rho(\omega)\exp(i\omega t)$, the longitudinal factor in the ih-RIDME signal is expressed as
\begin{equation}\label{eq:def_R}
    R(t; T_\text{mix}) = \int_{-\infty}^{+\infty} \mu_t(\omega, T_\text{mix})e^{-i\omega t}\mathrm{d}\omega\,.
\end{equation}
The hyperfine spectrum $\rho(\omega)$ can be well approximated by a zero-centred Gaussian function with standard deviation $\sigma$:
\begin{equation}
    \rho(\omega) = \frac{1}{\sqrt{2\pi}\sigma}\exp\left(-\frac{\omega^2}{2\sigma^2}\right)\,.
\end{equation}
With this assumption, the evolution is parametrised by two values: $D$ and $\sigma$. For locally homogeneous systems, where proton concentration is unambiguously defined, $\sigma$ is proportional to it\cite{kuzin2022diffusion}
\begin{equation}\label{eq:sigma_propto_CH}
    \sigma \propto c_H
\end{equation}
with the proportionality factor $\sigma/c_H = 0.0215~\si{\MHz\L\per\mol}$. The diffusion coefficient is analysed in a combination $D/\sigma^3$. This combination was experimentally found invariant under a homogeneous isotope dilution of the solvent mixture, resulting in a shape match of the ih-RIDME traces for different proton concentrations after the time axis transformation $t\rightarrow t \cdot c_{\textrm{H}}$, at all mixing times.\cite{kuzin2022diffusion} The parameter $D/\sigma^3$ correlates with an average number of close proton-proton contacts per single nucleus.\cite{kuzin2025ihridme} In water/glycerol mixtures, the coefficient $D/\sigma^3$ is slightly different for water protons and for glycerol protons. However, while the difference is detectable, it is so small that the ih-RIDME traces shapes for a given mixing time are very accurately scaling with the overall bulk proton concentration.\cite{kuzin2022diffusion}

Those nuclei contribute to the ih-RIDME decay whose homonuclear interaction with neighbours is strong enough to mix the hyperfine levels. This condition typically does not hold for the nearest few nuclei in a strong hyperfine field gradient. Such nuclear spin pairs are then `frozen' on the ih-RIDME time scale, and don't contribute to the ih-RIDME signal decay. The electron-proton distance with maximum contribution to ih-RIDME is $\approx 14$\,\AA\ in fully protonated solutions\cite{kuzin2022diffusion} and shifts to larger values as $\propto c_H^{-1/3}$. In proton-deuterium mixtures, considered in the present work, deuterium-induced spectral diffusion is also negligible.\cite{kuzin2022diffusion,kuzin2025ihridme} 

\subsection{ih-RIDME data analysis}
One typically measures a series of decays with different mixing times and analyses them globally. The normalized data $V(t; T_\text{mix})$ are described by a model\cite{kuzin2022diffusion,kuzin2024perturbation}
\begin{equation}\label{eq:V}
    V(t; T_\text{mix}) = R(t; T_\text{mix})\cdot F(t)
\end{equation}
The $R$-factor is given by Eq.~(\ref{eq:def_R}) and the $F$-factor arises since transverse evolution under homonuclear coupling is not included in the derivation of the $R$-factor.

For the signal in the five-pulse ih-RIDME experiment, the following approximations were found: \cite{kuzin2022diffusion,kuzin2025ihridme}
\begin{equation}\label{eq:R}
    R(t; T_\text{mix}) \approx \exp(-\alpha(T_\text{mix})\sigma^2t^2)
\end{equation}
with $\alpha(T_\text{mix}) = 1-\exp(-0.245(D/\sigma^3)T_\text{mix})$, and
\begin{equation}\label{eq:F}
    F(t) \approx \exp(-\beta\sigma^2t^2)\,,
\end{equation}
with $\beta \approx 0.13$ for five-pulse ih-RIDME. From Eqs.~(\ref{eq:V})-(\ref{eq:F}), a Gaussian shape of the ih-RIDME traces is expected, parameterised by mixing time. Proton density $\sigma$ determines the limiting curvature of the decay, and $D/\sigma^3$ regulates the build-up of this curvature.

We note that the ih-RIDME data can be analysed either directly using the numeric model (\ref{eq:sigma_propto_CH}-\ref{eq:F}), or by comparison with the reference data from a homogeneous sample of known proton concentration. The comparison is done by matching the traces' shape via global scaling of the time axis. The scaling coefficient is the ratio of the proton concentrations in two samples. Simultaneous analysis of multiple traces with varying mixing times ensures high accuracy in determining the proton concentration. If the ih-RIDME traces of the test sample cannot be precisely matched to the reference homogeneous sample, this is sufficient for concluding that the proton distribution around spin probes in the test sample is heterogeneous or that local proton concentration is distributed.\cite{kuzin2024quantification} Consequently, in a biphasic sample, the absence of accurate ih-RIDME traces scaling serves as proof that the spin probes are present in both phases. On the other hand, if all spin probes are within only one of the two phases, the shape congruence of ih-RIDME traces must apply, and the local proton concentration can be determined. While the proton concentration in an unknown homogeneous sample can also be determined with certain accuracy by other pulse EPR methods, such as Hahn echo decay\cite{jahn2022mechanism} or ESEEM measurements, the ih-RIDME technique is a tool to verify the homogeneity of solvent proton distribution.

\section{Experimental}

\subsection{Chemicals}
\ce{D2O} (Sigma Aldrich, 99.8 atom\% D), 2,2,6,6-tetramethylpiperidine-1-oxyl (TEMPO) (Sigma Aldrich, 99\%), \ce{H8}-glycerol (\ce{C3H5(OH)3}, Carl Roth, >99.7\%), \ce{D8}-glycerol (\ce{C3D5(OD)3}, Sigma Aldrich, >98 atom\% D) were used without further purification.

\subsection{Sample preparation}

Stock solution of TEMPO in \ce{D2O} ($c(\text{TEMPO}) = 1~\si{\milli M}$) was diluted in 10 times by \ce{D2O} or \ce{H2O} and mixed with protonated or perdeuterated glycerol according to Table~\ref{tab:sol_compos}. The mole fraction of glycerol in all solutions is 16-17\%, corresponding to the volume fraction 44-45\% and mass fraction 47-51\%. 30~\si{\micro\L} of each solution was transferred into a quartz EPR tube with an outer diameter in the range of 2.95-3.00~mm and shock-frozen in liquid nitrogen, ensuring the formation of a water-glycerol glass. Final concentration of TEMPO was 50~\si{\micro M}.

The EPR samples were characterized immediately after preparation (`fresh'), and then they were stored at -80~\si{\celsius} = 193~K (GFL Labor-Tiefk\"ultruhen). This value is above 158~K, which is the $T_g$ of the water-glycerol mixture.\cite{angarita2021revisiting} In order to test whether the spectroscopic changes from storing are reversible, we thawed and refroze the samples in EPR tubes. To this end, the tubes were turned upside down, and the sample volume was warmed up to room temperature while the quartz beyond the solution was kept in liquid nitrogen. This prevented the condensation of water vapours in the radical solution and the sliding of water droplets condensed above the solution upon storing it. The unfrozen solutions were kept in a liquid state for 5-10 seconds and shock-frozen in liquid nitrogen again.

\begin{table}[h!]
    \centering
    \begin{tabular}{ccccccc}\hline
        Sample & Water & Glycerol & $c_H$, M & Proton fraction, \% & $T_m,~\si{\micro\s}$ \\\hline
        \textbf{1} & H & H & 104 & 94.4 & 4.2 \\
        \textbf{2} & H & D & 56.4 & 51.1 & 6.4 \\
        \textbf{3} & D & H & 47.9 & 43.4 & 7.5 \\\hline
    \end{tabular}
    \caption{List of the solvents used in this study, their proton-deuteron composition, transverse ($T_m$) relaxation times of TEMPO in freshly frozen solutions, measured at 50~K. H and D correspond to protonated and perdeuterated chemicals. Proton fraction is calculated as $\frac{c_H}{c_H+c_D}$. Detailed description of solvent compositions can be found in ESI~S1.}
    \label{tab:sol_compos}
\end{table}

\subsection{EPR measurements}

All pulse EPR measurements were performed using a Q-band Bruker ElexSys spectrometer (MW frequency: 34–35 GHz), equipped with a home-built resonator designed for oversized 3 mm tubes.\cite{TSCHAGGELAR200981} Measurements were conducted at 50~K. Temperature stabilisation was established using an Oxford Instruments He-flow cryostat. The length of $\pi/2$- and $\pi$-pulses were $t_{\pi/2} = 12~\si\ns$ and $t_{\pi/2} = 24~\si\ns$, respectively. Hahn echo decay $(\pi/2)$–$t$–$(\pi)$–$t$–det ($t$ is incremented) was measured with a starting delay $t$ of 300 ns. The Hahn echo decay traces were used to determine the $1/e$ decay time, as well as for analysing the ESEEM contribution from deuterium nuclei.

Five-pulse RIDME measurements: the pulse sequence is shown in Fig.~\ref{fig:01}a, time delays were set to $d_1 = 0.4~\si{\us}$, $d_2 = 4.2~\si{\us}$. Values of mixing time were chosen as a geometric sequence $T_\text{mix} = 15 \times 2^n~\si{\us}\;(n = 0,...,5)$. In the RIDME measurements, the deuterium ESEEM-averaging protocol\cite{keller2016averaging} with 8 steps of 16 ns was used. All measurements were conducted at the maximum of the EPR spectrum (Fig.~\ref{fig:01}d).

\section{Results and Discussion}

The effect of `glass ageing' is known to EPR spectroscopists - the frozen solutions in water/glycerol when storing at -80~\si{\celsius} transform from the transparent glassy state to an opaque and white one, resembling snow. However, solutions shock-frozen directly to the liquid nitrogen temperature and handled without annealing, e.g., for EPR measurements at 50 K, or stored in liquid nitrogen, remain transparent. This appears to be associated with phase separation in this cold mixture occurring above the glass transition temperature. Murata and Tanaka investigated LLT in a water/glycerol mixture and determined the stability boundary for the water-rich liquid II phase approximately in the temperature range 150-200 K with the glycerol volume fraction changing between 30\% and 50\% as the temperature of the mixture decreases.\cite{murata2012liquid} Beyond this boundary, spontaneous phase separation into the liquid II and the glycerol-rich liquid I is followed by the crystallization of the liquid II.

The idea of our study is to deuterate either glycerol (solution 2) or water (solution 3). After LLT in these mixtures, the new phases have a different water-glycerol composition; consequently, they exhibit a different proton/deuterium ratio compared to a homogeneous solution. Changes in this ratio are probed by EPR, as decoherence time in Hahn echo and ih-RIDME experiments are sensitive to the local proton environment of a spin probe. After numerous tests, we concluded that the observed EPR changes are fully reversible, in the sense that refreezing the sample restores its transparency and resets the EPR properties to their initial values. Hence, one can exclude that the changes observed in aged samples are due to contamination by protonated water from the air, thus verifying the interpretation of the EPR parameters change as an internal transformation of the glass/cold liquid structure.

\subsection{Hahn echo decay and \texorpdfstring{\ce{^2H}}{2H}-ESEEM at Q band}

For nitroxide radicals in electron spin-diluted solutions (below 100~$\mu$M) at cryogenic temperatures, the rate of Hahn echo decay, as a function of the interpulse delay $t$, is strongly affected by the local proton environment.\cite{Eatons1998EPRDephasingbyProtons,ElMkami2014ProtDeut,Ward2010DistanceMeasurementsDeutProt} It was shown that decay traces can be simulated with a static spin Hamiltonian that includes hyperfine and homonuclear dipolar interaction.\cite{jeschke2023nuclear,jahn2022mechanism} At higher proton concentrations, the homonuclear interaction is stronger and the echo decays faster. The measured signal is the result of the deterministic evolution in the electron-nuclear spin system, where decay is due to the destructive interference of multiple nuclear modulation frequencies. For this reason, we prefer the term Hahn echo decay over transverse relaxation.

\begin{figure}[h!]
    \centering
    \includegraphics[width=1.0\linewidth]{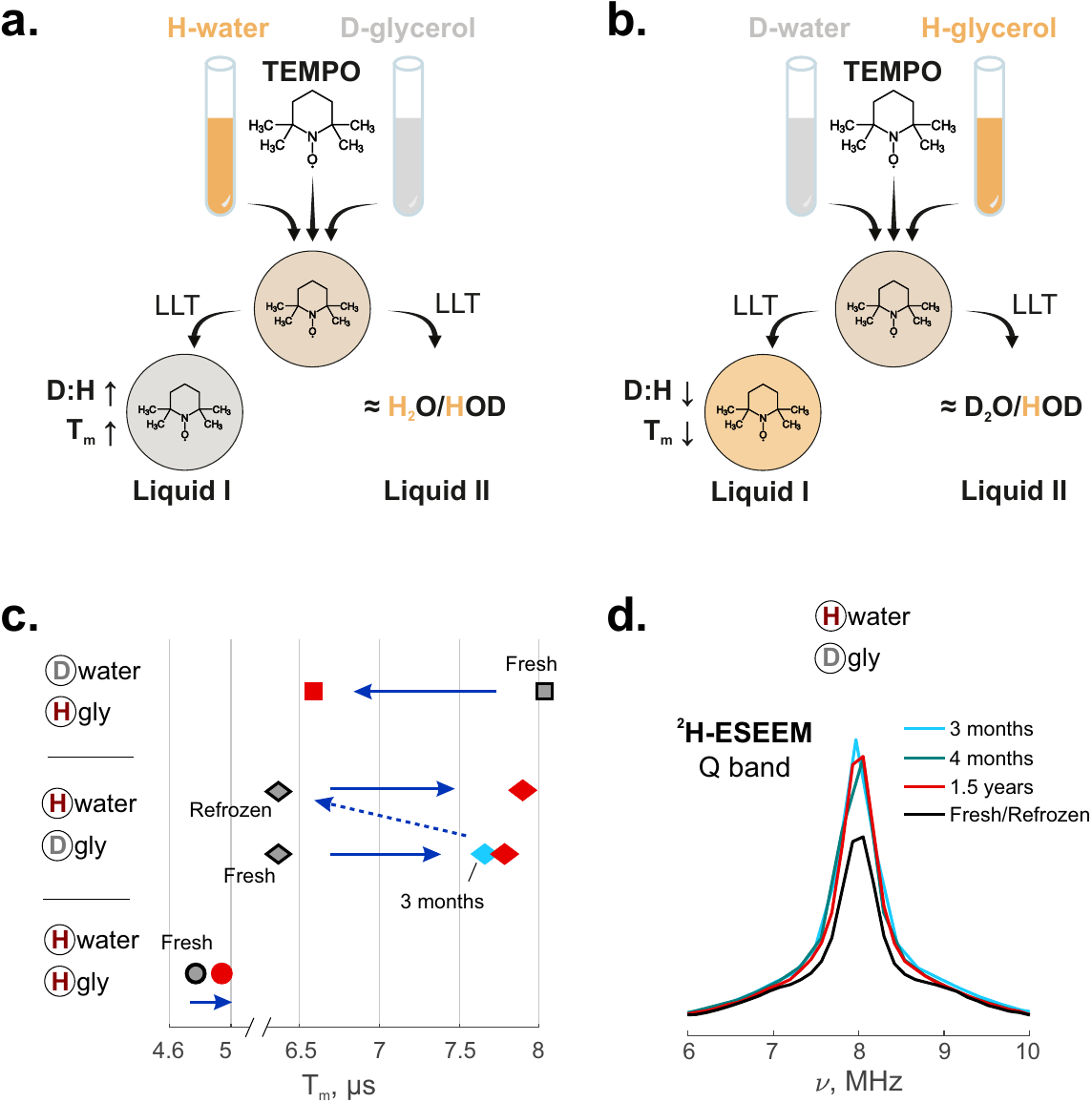}
    \caption{Spectroscopic changes of TEMPO in water/glycerol measured at Q band, 50~K. a,b. Schematic representation of the mixing and liquid-liquid transition process for the cases of \ce{H2O} and \ce{D_8}-glycerol (a) and \ce{D2O} and \ce{H_8}-glycerol (b). The protonation degree of a solution is pictured as a shade of orange; c. Changes of decoherence times (times of Hahn echo $1/e$ decay): grey symbols correspond to freshly frozen solutions, red data were measured after storing the samples for more than 1 year in a -80~\si{\celsius} freezer. Blue solid arrows indicate the direction of changes. A dashed arrow means refreezing the solution. The corresponding numeric values can be found in ESI Table~S2; d. Two-pulse ESEEM spectra (\ce{{}^2H} matrix peak, normalized by the intensity of the overtone).}
    \label{fig:glass:Te_ESEEM}
\end{figure}

The effect of sample annealing at -80~\si{\celsius} on the Hahn echo decay (Figure~\ref{fig:glass:Te_ESEEM}a) in the fully protonated medium is weak (4\% increase of the $T_m$ time). In contrast, we observe substantial changes for the partially deuterated solvents: 22\% increase for \ce{H_2O}+\ce{D_8}-glycerol and 18\% decrease for \ce{D_2O}+\ce{H_8}-glycerol. The observed changes are reversible as refreezing of the solutions precisely restores the `fresh' state. Within `thaw-freeze-store' cycles, the EPR changes are reproduced nearly quantitatively (ESI, Section~S2.3). For the shortest tested annealing time of 10 minutes, the LLT in the water-glycerol solutions was complete as judged from the Hahn echo data (ESI, Figure~S2). Long storage at -80~\si{\celsius} up to approx.\,1.5 years did not produce any further reliably detectable changes in the Hahn echo decay. As a reference, we stored a freshly frozen solution \textbf{2} in liquid nitrogen for up to 5 weeks and did not detect any measurable modifications to the glass structure (see ESI Section~S2.1 for details).

The opposite direction of changes in the two partially deuterated mixtures indicates that the local rearrangement is primarily determined by the chemical nature of the mixture component, water or glycerol, rather than by the isotope composition. We can qualitatively explain all three observations assuming depletion of water in the solvent shell around the nitroxide during ageing. (Figure~\ref{fig:glass:Te_ESEEM}a and b) Glycerol has five non-exchangeable hydrogen atoms in its molecule. Therefore, a glycerol-enriched environment would cause decreased local proton concentration with deuterated glycerol and increased proton concentration with protonated glycerol. The transverse relaxation time should increase in the former case and decrease in the latter case, as was observed. Only minor changes are expected in this scenario when both solvents are protonated. The observed slight increase of the phase memory time in solution 1 cannot be attributed to the remaining \ce{D_2O} from the TEMPO stock solution (5\%). The observed shift can be explained by differences in the shortest interproton distances in water and glycerol. In water, this is the intra-proton distance of 1.51~\AA, and in glycerol, it is the distance within a \ce{CH2} group, 1.78~\AA.\cite{glycerolGeometry} Hence, typical homonuclear coupling is weaker in protonated glycerol than in protonated water, which explains the prolongation of $T_m$.

Since solvents 2 and 3 are partially deuterated, we could also record the two-pulse ESEEM at Q band, where the signal from deuterons has sufficient modulation depth for the analysis and does not interfere with the proton signal (Figure~\ref{fig:glass:Te_ESEEM}b). In all cases, only weakly-coupled deuterons are seen around the Larmor frequency ($\nu_I(\ce{^2H}) \approx 8~\si{\mega\hertz}$ at Q band). The stored samples are characterised by a different modulation depth compared to the fresh ones, namely, it increases in solution 2 and decreases in solution 3 (see ESI~Section~S3). These results are consistent with the assumption that ageing depletes water in the vicinity of the nitroxide. 

\subsection{ih-RIDME analysis}

\begin{figure*}[h!]
    \centering
    \includegraphics[width=0.8\linewidth]{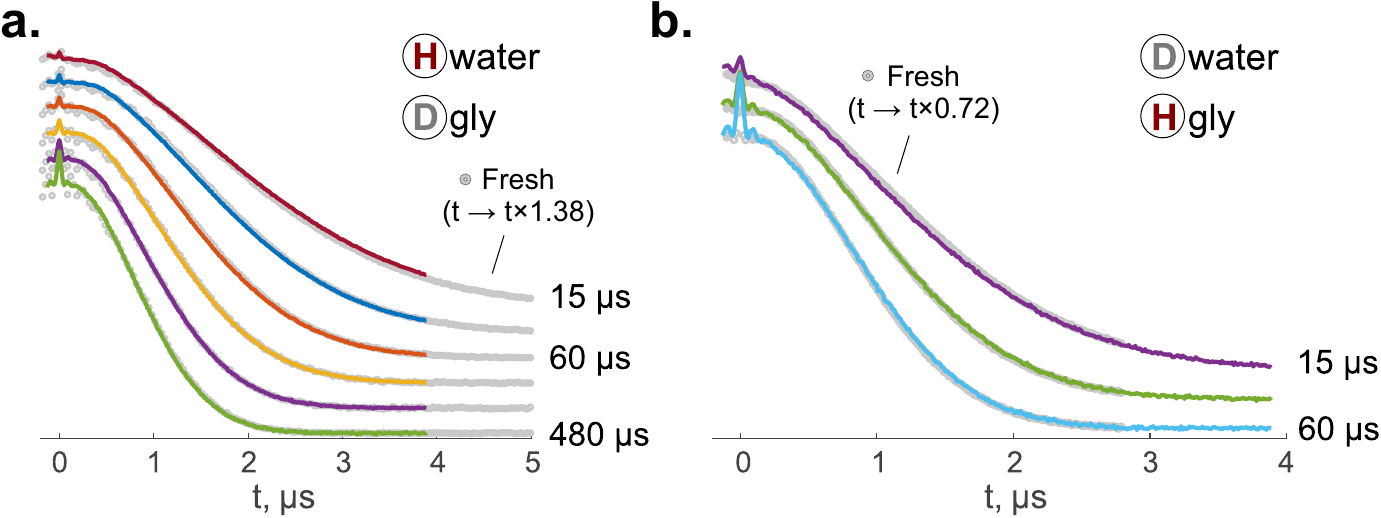}
    \caption{a,b. Overlaying of the normalized ih-RIDME traces of TEMPO in fresh (grey dots) and stored samples (coloured dots) for the case of (a) \ce{H2O} and \ce{D8}-glycerol and (b) \ce{D2O} and \ce{H8}-glycerol. The time axis of the fresh samples data is globally scaled, and the scaling factors $f$ are provided accordingly. The mixing times are specified next to the corresponding traces. The traces are shifted vertically for better visibility. Unscaled traces can be found in ESI~Section~S4.}
    \label{fig:ihRIDME}
\end{figure*}

The ih-RIDME data of fresh samples were published and analysed previously.\cite{kuzin2022diffusion} We showed that it can be fitted well assuming a single value for $\sigma$. We interpret this by referring to fresh solutions as locally homogeneous. In principle, there is no long-range molecular order in glassy mixtures; therefore, the distribution of water and glycerol molecules within the sensitivity range of ih-RIDME can fluctuate. However, the sensitivity region appears large enough (as estimated previously, this is roughly a sphere of about 2 nm radius for the given proton concentrations\cite{kuzin2022diffusion}). It contains many molecules, and thus large fluctuations in solvent composition have a negligible probability.

Similar to the Hahn echo decays, the ih-RIDME traces of TEMPO in stored glasses deviate from those of the fresh samples. We found that the ih-RIDME data of the fresh and stored samples can be matched by globally stretching the time axis. For the \ce{H_2O}+\ce{D_8}-glycerol sample (solution 2), the RIDME decays of the fresh sample are faster compared to the stored sample. By applying a factor of 1.38 to the ih-RIDME time axis of a fresh sample, we achieved the shape match for all traces simultaneously (Figure~\ref{fig:ihRIDME}a). The corresponding scaling factor for the \ce{D_2O}+\ce{H_8}-glycerol (solution 3) is 0.72 (Figure~\ref{fig:ihRIDME}b). The scaling values for the solutions 2 and 3 are approximately inverse of each other, which stems from the near equality of Hydrogen atoms in water and glycerol. Hence, the ih-RIDME signal change is in agreement with the Hahn echo decay data. This can be expected since both experiments rely on homonuclear coupling. However, the relative scaling in RIDME exceeds that of the relaxation data. This is evidence for the higher sensitivity of the RIDME experiment compared to the Hahn echo decay to such changes in the local nuclear environment. The global scaling symmetry of the datasets, similar to the case of homogeneous solutions in Ref.~\cite{kuzin2022diffusion}, determines that the variation of the $D/\sigma^3$ value is not significant. This is expected as the types of protons in the mixture are the same before and after molecular rearrangements. The previously determined weak differences in $D/\sigma^3$ between water and glycerol protons \cite{kuzin2022diffusion} are not resolved in these measurments. Consequently, the change of ih-RIDME data results solely from the change of $\sigma$, which, in turn, can be quantitatively interpreted as a change of local proton concentration around TEMPO.

To quantify the water/glycerol composition around TEMPO radicals after LLT, we derived the equations of proton balance, assuming the following points (illustrated in the Figure~\ref{fig:ihRIDME}c): (i) Chemical exchange of protons in water-based mixtures at room temperature is fast enough that it has fully equilibrated during the sample preparation;\cite{weinberg1955} (ii) Upon sample storing above the glass transition temperature, water diffusion is active, and it may form ice due to the instability of the liquid II phase against crystallization;\cite{murata2012liquid} (iii) Water ice contains neither glycerol nor TEMPO molecules. In the balance equation, we operate with the local protonation degree, $$p=\frac{N_p}{N_\text{hydr}}\,,$$ i.e. the ratio of protons ($N_p$) to the total amount of protons and deuterons in the sensitivity region of ih-RIDME. The protonation degree before LLT ($p_0$) is the same as the average solvent protonation degree calculated from the bulk isotope ratio. The total number of hydrogens in the mixture equals $N_\text{hydr} = h_w N_{w,0} + h_g N_{g,0}$, where $h_w = 2$ and $h_g = 8$ are the number of hydrogen atoms in water and glycerol, respectively, and $N_{w,0}$ and $N_{g,0}$ are the initial number water and glycerol molecules in the homogeneous mixture before LLT. In our model, we assume that water molecules leave the cold, viscous mixture to form liquid II, followed by spontaneous crystallisation, which consequently removes the hydrogen atoms. After hydrogen exchange with the hydroxyl groups of glycerol, the isotope composition of water is modified. We account for this by introducing the local protonation degree of water $p_w$. In equilibrium, this number is equal to the protonation degree of chemically exchangeable protons
\begin{equation}
    p_w = \frac{N_{p,ex}}{N_{ex}}\,,
\end{equation}
where $N_{p,ex}$ is the number of exchangeable protons and $N_{ex}$ is the number of all exchangeable hydrogens. The latter is computed identically for both mixtures as $h_{w,ex} N_{w,0} + h_{g, ex} N_{g,0}$ where $h_{w, ex} = 2, h_{g, ex} = 3$ are numbers of exchangeable protons in water and glycerol. In solution 2, protons originate from water; therefore, they are all chemically exchangeable and $N_{p,ex} = N_p = p_0 N_\text{hydr}$. This leads to the following formula for $p_w$ in solution 2
\begin{equation}
    p_{w, sol2} = \frac{p_0(h_w N_{w,0} + h_g N_{g,0})}{h_{w,ex} N_{w,0} + h_{g, ex} N_{g,0}} = \frac{p_0(h_w x_{w,0} + h_g x_{g,0})}{h_{w,ex} x_{w,0} + h_{g, ex} x_{g,0}}\,,
\end{equation}
where we introduced $x_{w,0}$ and $x_{g,0}$ - molar fractions of water and glycerol in the mixture. For solution 3, where protons stem from glycerol, we correct for non-exchangeable protons as $N_{p,ex} = N_p - (h_g - h_{g, ex})N_{g,0}$, hence obtain
\begin{equation}
    p_{w, sol3} = \frac{p_0(h_w x_{w,0} + h_g x_{g,0})-(h_g - h_{g, ex})x_{g,0}}{h_{w,ex} x_{w,0} + h_{g, ex} x_{g,0}}\,.
\end{equation}

\begin{figure}[h!]
    \centering
    \includegraphics[width=1\linewidth]{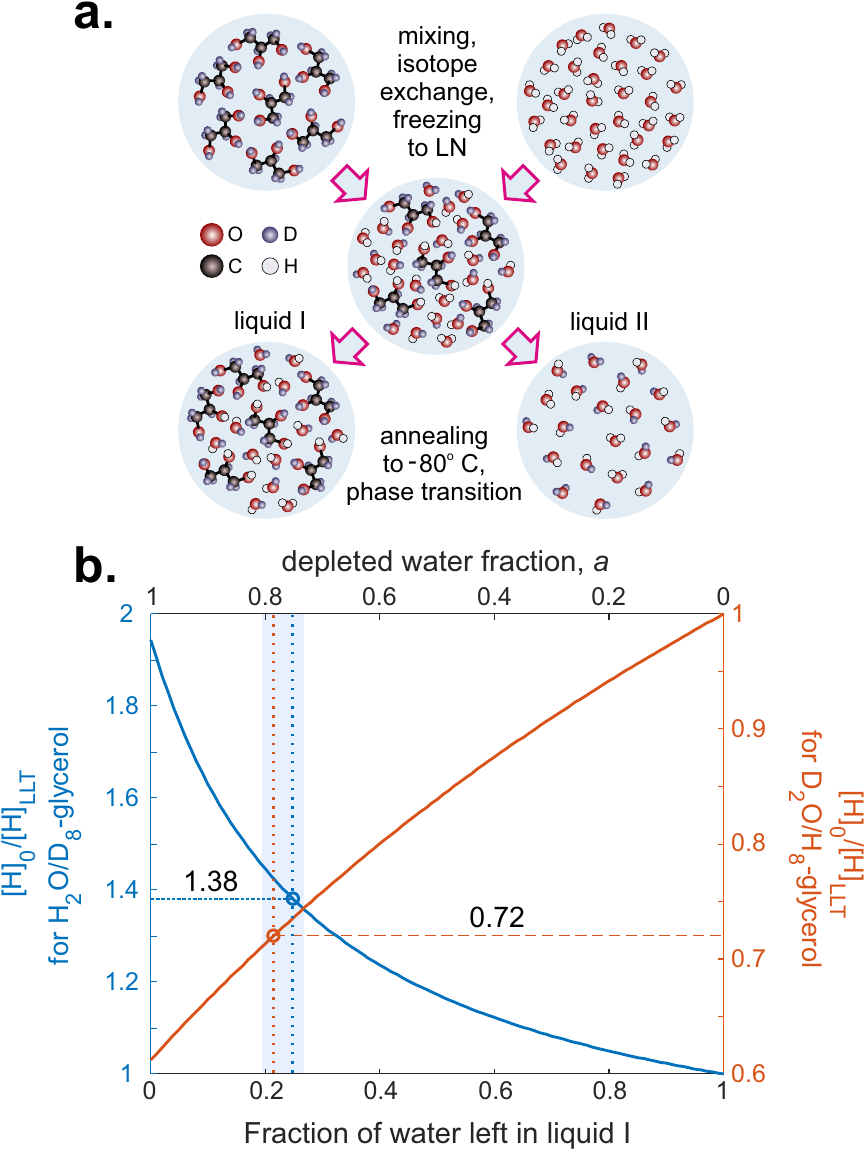}
    \caption{a. Schematic molecular level representation of water and glycerol mixing, isotope exchange and liquid-liquid transition for the case of \ce{H2O} and \ce{D8}-glycerol. b. Ratio of the initial protons concentration in the fresh sample ($[\textrm{H}]_0$) and the proton concentration in the liquid I after isotope exchange and liquid-liquid transition ($[\textrm{H}]_{\textrm{LLT}}$) for the case of \ce{H2O} and \ce{D8}-glycerol (blue curve, left vertical axis) and \ce{D2O} and \ce{H8}-glycerol (red curve, right vertical axis). Circles indicate the corresponding stretching factors for the ih-RIDME data, while the vertical band indicates the possible scatter of the molar fraction values due to measurement uncertainty and possible temperature variations during annealing.}
    \label{fig:fig04}
\end{figure}

We describe the effect of LLT by a parameter $0\leq a\leq 1$, the relative fraction of water molecules that leaves the solvent shell. In these terms, the observed local protonation degree after water depletion is
\begin{equation}\label{eq:applSpecD:glass:proton_balance}
    p_\text{LLT} = \frac{p_0 N_\text{hydr} - p_w h_w a N_{w,0}}{N_\text{hydr} -  h_w a N_{w,0}} = \frac{p_0 (h_w x_{w,0} + h_g x_{g,0}) - p_w h_w a x_{w,0}}{(h_w x_{w,0} + h_g x_{g,0}) -  h_w a x_{w,0}}
\end{equation}
where $a N_{w,0}$ is the average number of removed water molecules. Also, in this expression, we assumed that water depletion is an isotope-non-selective process. The ratio $p_0/p_\text{LLT}$, calculated for solutions 2 and 3, is plotted in Figure~\ref{fig:fig04}b as a function of $a$. For solution 2, this curve is above 1, meaning that LLT always decreases local proton concentration. For solution 3, the situation is opposite, so the curve lies below 1.

With the direct proportionality relation between $\sigma$ and $c_H$ in ih-RIDME (see Eq.~(\ref{eq:sigma_propto_CH})), we calculated for the solution 2 $p_\text{LLT}=p_0/1.38$. Further, the Eq.~(\ref{eq:applSpecD:glass:proton_balance}) was solved for $a$ which yielded $a = 75.1\pm0.8\%$. For solution 3, for which $p_\text{LLT}=p_0/0.72$, we obtained $a=78.6\pm2.1\%$ (see Table~\ref{tab:applSpecD:glass_calculation}). These values are close for both solutions, albeit not the same, which may indicate instability of the temperature regime at long incubations. At the same time, this might also be due to glass ageing being weakly sensitive to the isotope distribution between water and glycerol. Overall, one can conclude that the deuteration creates an efficient contrast for spectral diffusion and relaxation with a minimal perturbation of the LLT processes.

\begin{table*}[h!]
    \centering
    \begin{tabular}{ccccccc}\hline
        Solution & $x_0$ & $p_0$ & $p_w$ & $f$ & $p_\text{LLT}$ &  $a, \%$ \\ \hline
        2 & $0.840$ & $0.511$ & $0.700$ & $1.38\pm0.01$ & $0.370\pm0.002$ & $75.1\pm0.8$\\
        3 & $0.839$ & $0.434$ & $0.223$ & $0.72\pm0.01$ & $0.603\pm0.008$ & $78.6\pm2.1$\\ \hline
    \end{tabular}
    \caption{Numerical details on the water depletion analysis in solutions 2 and 3. $x_0$ is the bulk water molar fraction, $p_0$ is the bulk protonation degree, $p_w$ is the water protonation degree after isotope exchange with glycerol, $f$ is the ih-RIDME time axis stretching factor (see Figure~\ref{fig:ihRIDME}a-b) after LLT, $p_\text{LLT}$ is the local protonation degree probed by TEMPO after LLT, $a$ is the calculated water depletion degree from liquid I. Uncertainties of $p_\text{LLT}$ and $a$ were propagated from the uncertainty of $f$.}
    \label{tab:applSpecD:glass_calculation}
\end{table*}

\subsection{Discussion}

In Ref.~\cite{jahn2022mechanism}, Hahn echo decay for perdeuterated TEMPO in water/glycerol mixtures was measured in a full deuteration range, and the relation $T_m \propto c_\mathrm{H}^{-0.65}$ was proposed. Relative changes in $T_m$ detected by us and recalculated to proton concentrations with this power law are in good agreement with ih-RIDME. The corresponding changes of $c_H$ after LLT are $1.36$ for solution 2 ($1.38$ in ih-RIDME) and $0.74$ for solution 3 ($0.72$ in ih-RIDME). We note that exact shape congruence for Hahn echo decays is not expected because (i) the TEMPO methyl tunneling contribution\cite{eggeling2023quantifying} does not scale with bulk proton concentration, (ii) the stretched exponential parameter $\xi$ also changes with proton concentration.\cite{jahn2022mechanism} Consequently, a quantitative analysis of the proton bath contribution in Hahn echo decays requires the explicit fitting of these contributions with model functions.

\begin{table}[h!]
    \centering
    \begin{tabular}{ccccc}\hline
        \multicolumn{5}{c}{Before LLT} \\\hline
        Solution & $x_w$ & $x_g$ & $\phi_w$ & $\phi_g$\\
        2 & $84.0$ & $16.0$ & $56.5$ & $43.5$ \\
        3 & $83.9$ & $16.1$ & $56.3$ & $43.7$ \\
        \hline
        \multicolumn{5}{c}{After LLT} \\\hline
        2 & $56.7\pm0.8$ & $43.3\mp0.8$ & $24.4\pm0.6$ & $75.6\mp0.6$ \\
        3 & $52.3\pm1.2$ & $47.7\mp1.2$ & $21.6\pm1.6$ & $78.4\mp1.6$ \\
        \hline
    \end{tabular}
    \caption{Comparison of water/glycerol composition around spin probe before and after LLT based on ih-RIDME data. $x$ and $\phi$ represent molar and volume fractions of water $w$, respectively, glycerol $g$. All values are expressed as percentages.}
    \label{tab:before_after}
\end{table}

We can safely conclude from the ih-RIDME data that the statistical majority of the spin probes stays in the glycerol-rich liquid I phase. This follows from the precise shape congruence of the ih-RIDME traces. Should spin probes populate both phases and be present in two very different environments in terms of proton concentration, global matching by traces scaling would not be possible. Besides, the composition of the liquid I must correspond to thermodynamic equilibrium, which follows from two observations. Firstly, changes of Hahn echo and ih-RIDME traces upon LLT are reproducible (see data with solution 2 in Figure~\ref{fig:glass:Te_ESEEM}c). Secondly, RIDME data indicate that the proton concentration in liquid I is uniform throughout the entire sample.

Water depletion occurs within at least 2-3~\si{\nm} from the spin probe, corresponding to the estimated sensitivity range of the ih-RIDME method.\cite{kuzin2022diffusion} At the spin concentration of 50~\si{\micro M} with the homogeneous spin distribution, the average inter-probe distance is 32~\si\nm. Ref.~\cite{murata2012liquid} evidences that in frozen water-glycerol mixtures, the changes during such phase separation processes take place on a micrometre scale. However, the TEMPO molecules do not sense the agglomeration of water. It is thus likely that the TEMPO molecules are pushed from the nucleation and crystallisation centres together with the 2-3~\si\nm\, glycerol-rich solvent sphere. This may be related to the high viscosity of glycerol in the liquid state. It forms a branched net of strong hydrogen bonds around TEMPO molecules, keeping them caged in the rigid frame. Alternatively, the viscosity of the cold water/glycerol mixture at the storage temperature still permits that TEMPO molecules can diffuse and form a homogeneous solution in the glycerol-rich phase.\cite{Feldman2015wgice}

To discuss the overall ageing picture of frozen TEMPO solutions, we can focus on two main types of transformation. The first one, the nearest to the spin centre, is the breaking and building of H-bonds with the NO-group of the nitroxide radical. This has been efficiently probed previously via ESEEM and electron-nuclear double resonance (ENDOR) spectroscopies.\cite{bordignon2015_glasses} Such a pathway is the most sensitive to solvent deuteration due to the isotope effect on the hydrogen bond strength. Interestingly, direct contact of TEMPO with water molecules and formation of a new hydrogen bond were demonstrated upon annealing. This appears counterintuitive at first glance since, in fact, our EPR data demonstrate that water depletion takes place in the vicinity of TEMPO molecules upon annealing, which is the second type of transformation. We can argue that the molecular reorganisation and LLT at low temperature are due to a shift in the balance between enthalpic and entropic contributions to the solvent and solute chemical potentials. It is likely that the weak energy differences in the specific interaction between water and glycerol molecules favour their separation at lower temperatures, while they cannot do this at ambient temperature due to the stronger impact of the entropic terms.

Murata and Tanaka reported formation of only small water crystals, of approximately 11 nm size, after LLT in water glycerol mixtures below 205 K, while above this temperature macroscopic ice growth has been observed.\cite{murata2012liquid} Since the incubation temperature in our experiments is only eight to ten degrees below this limit and also our incubation conditions admit some temperature variations in the range of few Kelvin, we cannot exclude bulk ice formation in our stored/annealed samples. From the fraction of water leaving the vicinity of spin probes (75\%), we conclude that about 42\% of the volume is occupied by the water-rich phase after the LLT. Thus, in our experimental conditions, it must turn to water crystals, leaving the remaining 58\% of the sample volume in a deeply cooled liquid I state. 

The separation into two deeply cooled liquid phases in the water glycerol mixtures is a bulk effect and is therefore most likely solute-independent, at least at the given \si{\micro M}-range of solute (TEMPO) concentrations. The overall vibrational modes distribution in the new glass formed after ageing should not be very different from the original one in the homogeneous water-glycerol mixture, because we observe no change in the longitudinal relaxation time $T_1$ of TEMPO radicals. An interesting experiment to disentangle different contributions would be a comparison of the glass evolution kinetics from the EPR data, IR and terahertz spectroscopy, and the phase-sensitive microscopy to correlate these changes on different length and time scales. To provide sufficient time resolution, such experiments would need to be conducted at lower temperatures, closer to the glass transition temperatures of the homogeneous mixture.

Having quantified the molar composition of the liquid I phase after LLT, we can also explain the apparent mismatch between the glass transition temperature for annealed and non-annealed samples.\cite{bordignon2015_glasses,angarita2021revisiting} In the CW EPR power saturation measurements for annealed and not annealed water/glycerol mixtures, a shift of the glass transition temperature was determined, which is in line with the LLT described by Murata and Tanaka and quantified here. Additionally, the time scale of the CW EPR measurements is quite long (tens of minutes to hours), which likely resulted in LLT occurring during the measurements above the glass transition temperature. The LLT is the most realistic explanation for the rise of the saturation parameter $P_{1/2}$ just above the glass transition temperature, resulting in a discontinuity in the saturation curve. The measurements on annealed samples indicated a higher glass transition temperature compared to the freshly frozen samples, consistent with the data for mixtures with a higher glycerol molar fraction appearing after LLT. Note that the glass transition temperature of about 181~K after annealing is consistent with the liquid I molar composition reported in the present work. Thus, we can conclude that an accurate analysis of the LLT in water/glycerol mixtures and the phases quantification by ih-RIDME can also explain the seemingly inconsistent data in previously published reports.

Another question that appears interesting in this connection is the isotopic composition of the water ice and amorphous water phase formed in the frozen mixture \ce{H_2O}-\ce{D_2O}, e.g. in the agarose-gel-stabilised samples without glycerol.\cite{emmanouilidis2021nmr} On the one hand, the isotope effect on the strength of hydrogen bonds is well known. On the other hand, it is not well known whether this effect is strong enough to enable partial isotope separation upon fast freezing in supercooled isopentane, used in the cited work. As described above, in our RIDME-based ageing experiments with water-glycerol mixtures, we obtained slightly different (relative difference by about 5\%) composition of the bulk phase around TEMPO radicals in the case of H-glycerol and D-water as compared to the reverted case of D-glycerol and H-water. While this might be within experimental uncertainty, e.g. due to the variations of the temperature during annealing, it also leaves the possibility of a weak isotope effect being present in these experiments. We can thus imagine that, due to the isotope effect, the process of ice formation, occurring at low temperatures and on a slow enough time scale, might turn out to be partially isotope-selective. It would be interesting to verify or disprove this assumption in future experiments.

The first solvation shell of TEMPO molecules, addressable in ENDOR and ESEEM experiments, should mostly fall into the insensitive short distance range in ih-RIDME. However, having demonstrated here the usefulness of the ih-RIDME technique for determining the local compositions in deeply frozen water/glycerol mixtures, we can propose that ih-RIDME experiments on spin-labelled biomacromolecules might also appear useful for investigating the solvation shell peculiarities of protein residues in the vicinity of the spin-labelled site. In our lab, we have already demonstrated the applicability of ih-RIDME for studying polysaccharides\cite{kuzin2024quantification}, and we are currently performing ih-RIDME tests on site-specifically spin-labelled protein molecules.

\section*{Conclusions}

In this work, we demonstrated how pulse EPR can be utilised to investigate the liquid-liquid transition in cold water/glycerol mixtures. Thanks to the specific deuteration of water or glycerol, the local proton concentration in the vicinity of the TEMPO radical changes upon LLT, as probed by Hahn echo and ih-RIDME experiments. The ih-RIDME traces show global shape congruence at all mixing times, allowing us to conclude that, firstly, TEMPO is located in the glycerol-rich liquid I, and secondly, the proton distribution in liquid I is homogeneous. By quantifying local proton concentration with ih-RIDME, we calculated the volume ratio of liquid I to liquid II 52\%:48\%. It is equivalent to say that upon LLT, $75.1\pm0.8\%$ (in \ce{H2O}/\ce{d_8}-glycerol) and $78.6\pm2.1\%$ (in \ce{D2O}/\ce{h_8}-glycerol) of water forms a separate phase and crystallises.

New data help bring together the previously published results on glass transition temperatures, hydrogen bonding of spin probes and phase stability of water/glycerol solutions. The ih-RIDME technique proves to be a reliable and precise tool for determining local molar composition in binary mixtures. It can therefore be proposed also for the tempting studies on the solvation of biomacromolecules.

\section*{Acknowledgements}
The authors would like to acknowledge Dr.~Daniel Klose for bringing our attention to the work of Murata and Tanaka\cite{murata2012liquid}, and Prof.~Gunnar Jeschke for numerous fruitful discussions. Swiss National Science Foundation is acknowledged (Grant No. 200020\_188467).

\newpage

\section{Supplementary Information}
\subsection{Detailed mixture composition}

\begin{table}[h!]
    \centering
    \begin{tabular}{|c|ccc|ccc|}\hline
         \multicolumn{7}{|c|}{\textbf{Mass fraction}} \\\hline
         & \ce{D2O} & \ce{H2O} & water & \ce{d8-gly} & \ce{h8-gly} & glycerol \\
         sol1 & 5.5 & 43.7 & 49.2 & 0 & 50.8 & 50.8 \\
         sol2 & 5.4 & 43.5 & 48.9 & 51.1 & 0 & 51.1 \\
         sol3 & 53.1 & 0 & 53.1 & 0 & 46.9 & 46.9 \\
         \hline\hline
         \multicolumn{7}{|c|}{\textbf{Volume fraction}} \\\hline
         & \ce{D2O} & \ce{H2O} & water & \ce{d8-gly} & \ce{h8-gly} & glycerol \\
         sol1 & 5.6 & 49.2 & 54.7 & 0 & 45.3 & 45.3 \\
         sol2 & 5.7 & 50.8 & 56.5 & 43.5 & 0 & 43.5 \\
         sol3 & 56.3 & 0 & 56.3 & 0 & 43.7 & 43.7 \\
         \hline\hline
         \multicolumn{7}{|c|}{\textbf{Molar fraction}} \\\hline
         & \ce{D2O} & \ce{H2O} & water & \ce{d8-gly} & \ce{h8-gly} & glycerol \\
         sol1 & 8.4 & 74.6 & 83.1 & 0 & 16.9 & 16.9 \\
         sol2 & 8.5 & 75.5 & 84.0 & 16.0 & 0 & 16.0 \\
         sol3 & 83.9 & 0 & 83.9 & 0 & 16.1 & 16.1 \\
         \hline
    \end{tabular}
    \caption{Detailed description of water/glycerol mixture composition represented in mass, volume and molar fractions. Columns `water' and `glycerol' contain the sum values of all water and glycerol forms, respectively. All values are expressed as percentages.}
    \label{tab:si:composition}
\end{table}

\newpage
\subsection{Hahn echo decay data}

\subsubsection{Storing in liquid nitrogen}

\begin{figure}[h!]
    \centering
    \includegraphics[width=0.5\linewidth]{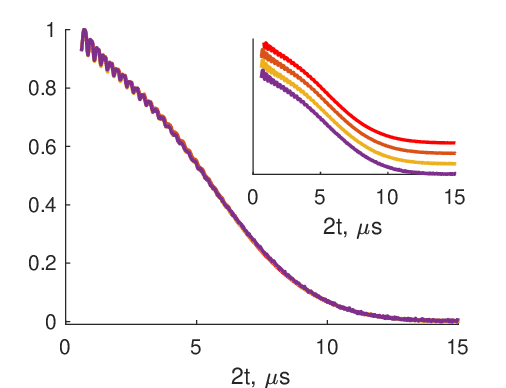}
    \caption{Superimposed and normalized Hahn echo decay traces of TEMPO in solution 2 stored in liquid nitrogen. The inset shows the same trace vertically shifted for better visibility. Storing conditions from top to bottom: freshly frozen, 1 week, 3 weeks, 5 weeks. Measurements were performed at Q band, 50 K.}
    \label{fig:SI:LN2}
\end{figure}

We demonstrated that storing water/glycerol mixtures deeply below the glass transition temperature preserves their EPR properties. Figure~\ref{fig:SI:LN2} shows Hahn echo decay traces $[\pi/2 - t - \pi - t - \text{echo}]$ of TEMPO in freshly frozen solution 2 (\ce{H2O}/\ce{d8}-glycerol, red line) and after storing it in liquid nitrogen (77\,K) for 1, 3 and 5 weeks. Traces perfectly overlap, which demonstrates the stability of the water/glycerol glass under these conditions.

\subsubsection{Annealing in dry ice}

\begin{figure}[h!]
    \centering
    \includegraphics[width=0.5\linewidth]{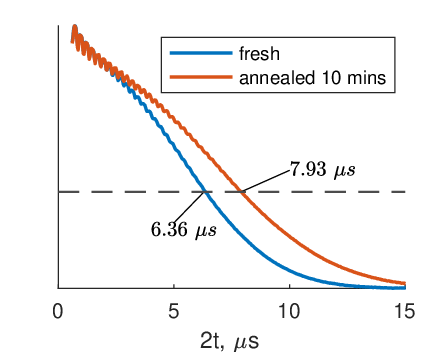}
    \caption{Normalised Hahn echo decays of TEMPO in solution 2 (\ce{H2O}/\ce{d8}-glycerol) as freshly frozen (blue) and annealed in dry ice for 10 mins (orange). The dashed line shows $1/e$ value. Measurements were performed at Q band at a temperature of 50~K.}
    \label{fig:SI:dry_ice}
\end{figure}

An experiment with short annealing in dry ice was performed as follows. An EPR tube with TEMPO in solution~2 (\ce{H2O}/\ce{d8}-glycerol) was freshly frozen in liquid nitrogen, and the Hahn echo decay was measured (Q band, 50~K - blue line in Figure~\ref{fig:SI:dry_ice}). The decay time measured at $1/e$ of initial intensity amounted to 6.36~\si{\us}, which reproduces other measurements with fresh samples. After that, the tube was quickly immersed in the crushed dry ice. At this point, the sample in it was transparent. The top of the tube was firmly covered with Teflon tape to prevent water vapour condensation. After ten minutes, the tube, in which the content turned white and non-transparent, was rapidly transferred to liquid nitrogen for quenching further changes. The Hahn echo decay measurement was repeated (orange line in Figure~\ref{fig:SI:dry_ice}), and the characteristic decay time increased to 7.93~\si{\us}. This value aligns with data obtained in other annealing experiments, which evidences that the LLT processes have completed.

\newpage
\subsubsection{Storing at $-80$~\si{\celsius}}

\begin{figure}[h!]
    \centering
    \includegraphics[width=0.5\linewidth]{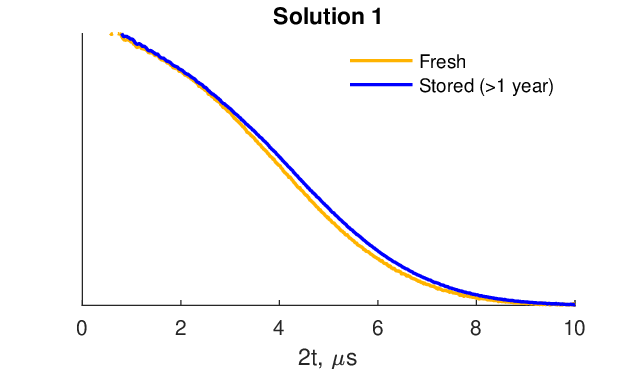}\\[1ex]
    \includegraphics[width=0.5\linewidth]{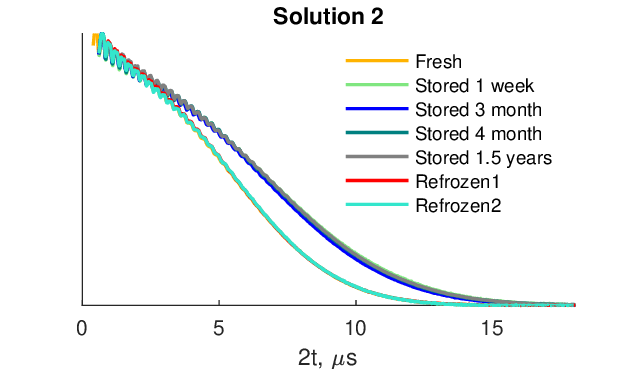}\\[1ex]
    \includegraphics[width=0.5\linewidth]{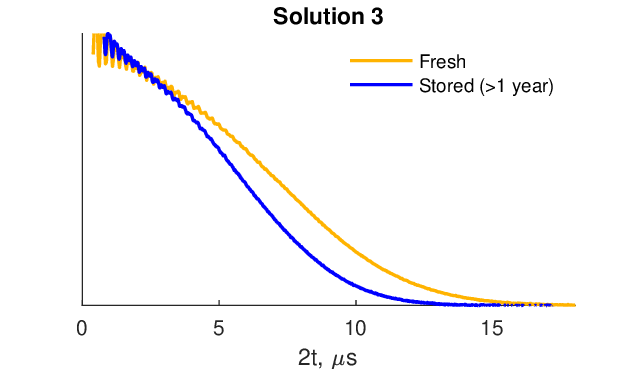}
    \caption{Normalised Hahn echo decay traces of TEMPO in solutions 1-3: freshly frozen, stored at -80~\si{\celsius} and refrozen (see main text for details). The traces correspond to the values in Table~\ref{tab:SI:Tm}.}
    \label{fig:SI:all_HahnEchoDecays}
\end{figure}

In Table~\ref{tab:SI:Tm}, we collect values of $T_m$ shown in Figure~2c in the main text. The corresponding time traces are shown in Figure~\ref{fig:SI:all_HahnEchoDecays}.

\begin{table}[h!]
    \centering
    \begin{tabular}{ccc}\hline
        Solution & State & $T_m$, \si{\us} \\\hline
        1 & Fresh & $4.75$ \\
         & Stored ($>1$ year) & $4.94$ \\\hline
        2 & Fresh & $6.35$ \\
         & Stored ($>1$ year) & $7.78$ \\
         & Refrozen & $6.37$ \\
         & Stored (3 months) & $7.67$ \\
         & Stored (4 months) & $7.78$ \\
         & Refrozen & $6.39$ \\
         & Stored (1 week) & $7.82$ \\\hline
        3 & Fresh & $8.05$ \\
         & Stored ($>1$ year) & $6.59$ \\\hline
    \end{tabular}
    \caption{Characteristic Hahn echo decay times for TEMPO in solutions 1-3. `Stored' refers to storing the EPR tube with solution at -80~\si{\celsius} in a commercial freezer. Data were measured at Q band at 50~K.}
    \label{tab:SI:Tm}
\end{table}

\newpage
$ $
\newpage
\subsection{Two-pulse ESEEM}

\begin{figure}[h!]
    \centering
    \includegraphics[width=0.5\linewidth]{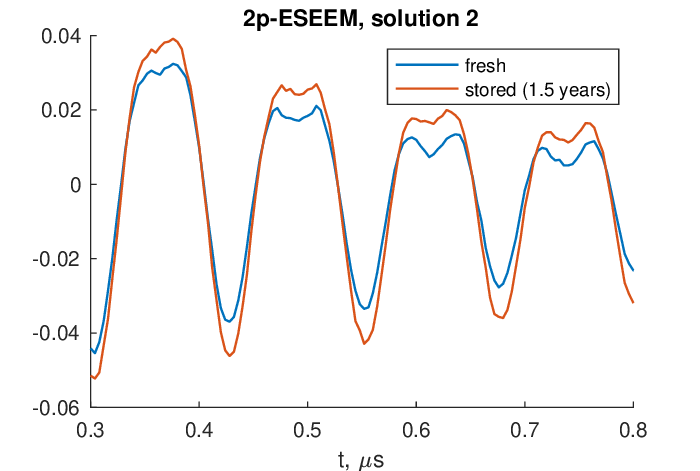}\\
    \vspace{1em}
    \includegraphics[width=0.5\linewidth]{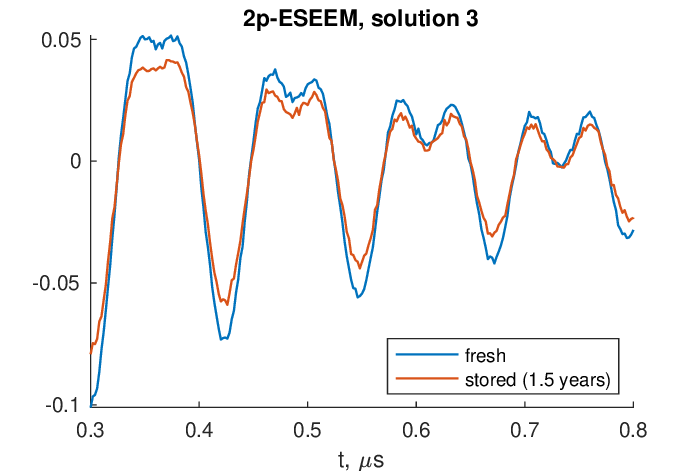}
    \caption{Fragments of two-pulse ESEEM traces for TEMPO in solution 2 (top) and solution 3 (bottom) as freshly frozen (blue line) and after long incubation at $-80~\si{\celsius}$ (orange line). Traces were normalised by the maximum point and with the following subtraction of a polynomial background of order 4. Data are acquired at Q band, 50 K, as $\pi/2-t-\pi-t-\text{echo}$ with initial delay $t = 0.3~\si{\us}$ and step 4~\si{\ns}.}
    \label{fig:SI:2p_ESEEM_TD}
\end{figure}

Figure~\ref{fig:SI:2p_ESEEM_TD} contains background-corrected two-pulse ESEEM traces of TEMPO in solutions 2 (\ce{H2O}/\ce{d8}-glycerol) and 3 (\ce{D2O}/\ce{h8}-glycerol) in the fresh state and after long incubation at $-80~\si{\celsius}$. Distinct oscillations with a period of $\approx 128~\si{\ns}$ are observed, representing weakly coupled deuterium at Q band. Primary data were normalised by the maximum point; therefore, the amplitude of oscillations corresponds to the ESEEM modulation depth. For weakly coupled nuclei, the following approximation for the total modulation depth applies
\begin{equation}
    k_\text{tot} \approx \sum_{i} k_i\,
\end{equation}
where $k_i$ is the single-nucleus ESEEM modulation depth. Consequently, $k_\text{tot}$ can be a measure of the deuterium number in the vicinity of a spin probe. We observe opposite changes in solutions 2 and 3: after the liquid-liquid transition (LLT), the modulation amplitude in solution 2 increases, while in solution 3, it decreases. We interpret this as the parallel changes of local deuterium concentration. These data are compatible with the assumption that TEMPO is located in liquid I after the LLT (see Figure~2a-b in the main text).

\begin{figure}[h!]
    \centering
    \includegraphics[width=0.75\linewidth]{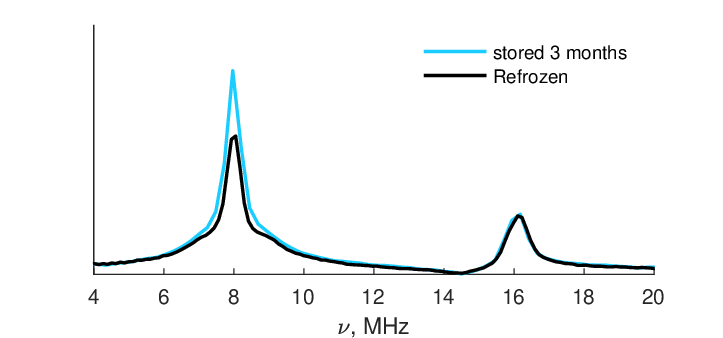}
    \caption{Two-pulse ESEEM spectra of fresh (black line) and stored (light blue line) solution 2. The frequency range includes the matrix peak at the deuterium Larmor frequency and the combination peak. The spectra are normalised by the intensity of the combination peak.}
    \label{fig:SI:2p_ESEEM}
\end{figure}

\newpage
\subsection{ih-RIDME traces}

\begin{figure}[h!]
    \centering
    \includegraphics[width=0.7\linewidth]{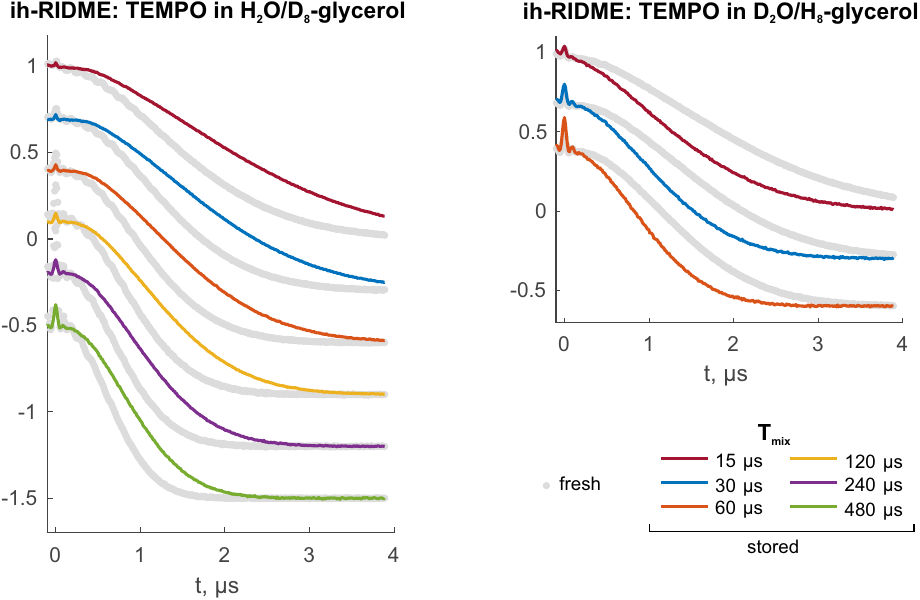}
    \caption{Experimental ih-RIDME traces of TEMPO in solution 2 (right) and 3 (left). Grey traces are measured with freshly frozen solutions. Coloured traces correspond to the stored samples. Traces are shifted vertically for better visibility. The legend with mixing times in the bottom right corner applies to both charts.}
    \label{fig:SI:ihridme}
\end{figure}

Figure~\ref{fig:SI:ihridme} shows ih-RIDME traces of TEMPO in solution 2 (\ce{H2O}/\ce{d8}-glycerol, left panel) and solution 3 (\ce{D2O}/\ce{h8}-glycerol, right panel). Grey dots correspond to fresh samples and coloured lines represent samples after LLT (induced by storing at -80~\si{\celsius}). In solution 2, the decays are slower after LLT, and in solution 3, they are faster. This evidence is compatible with the assumption that TEMPO is located in liquid I after the LLT. The datasets for solution 2 can be matched for all mixing times after multiplying the time axis of the fresh state by a factor of 1.38. The same applies to solution 3, where the stretching factor amounts to 0.72 (see Figure 3a-b in the main text).

\bibliography{references} %You need to replace "rsc" on this line with the name of your .bib file

\providecommand*{\mcitethebibliography}{\thebibliography}
\csname @ifundefined\endcsname{endmcitethebibliography}
{\let\endmcitethebibliography\endthebibliography}{}
\begin{mcitethebibliography}{41}
\providecommand*{\natexlab}[1]{#1}
\providecommand*{\mciteSetBstSublistMode}[1]{}
\providecommand*{\mciteSetBstMaxWidthForm}[2]{}
\providecommand*{\mciteBstWouldAddEndPuncttrue}
  {\def\EndOfBibitem{\unskip.}}
\providecommand*{\mciteBstWouldAddEndPunctfalse}
  {\let\EndOfBibitem\relax}
\providecommand*{\mciteSetBstMidEndSepPunct}[3]{}
\providecommand*{\mciteSetBstSublistLabelBeginEnd}[3]{}
\providecommand*{\EndOfBibitem}{}
\mciteSetBstSublistMode{f}
\mciteSetBstMaxWidthForm{subitem}
{(\emph{\alph{mcitesubitemcount}})}
\mciteSetBstSublistLabelBeginEnd{\mcitemaxwidthsubitemform\space}
{\relax}{\relax}

\bibitem[Bauer \emph{et~al.}(2017)Bauer, Dotta, Balacescu, Gath, Hunkeler,
  Bockmann, and Meier]{Bauer2017LTSSNMR}
T.~Bauer, C.~Dotta, L.~Balacescu, J.~Gath, A.~Hunkeler, A.~Bockmann and B.~H.
  Meier, \emph{J. Biomol. NMR}, 2017, \textbf{67}, 51--61\relax
\mciteBstWouldAddEndPuncttrue
\mciteSetBstMidEndSepPunct{\mcitedefaultmidpunct}
{\mcitedefaultendpunct}{\mcitedefaultseppunct}\relax
\EndOfBibitem
\bibitem[Callon \emph{et~al.}(2022)Callon, Malär, Lecoq, Dujardin, Fogeron,
  Wang, Schledorn, Bauer, Nassal, Böckmann, and Meier]{Meier2022FastMASHepVB}
M.~Callon, A.~A. Malär, L.~Lecoq, M.~Dujardin, M.-L. Fogeron, S.~Wang,
  M.~Schledorn, T.~Bauer, M.~Nassal, A.~Böckmann and B.~H. Meier, \emph{Angew.
  Chem. Int. Ed.}, 2022, \textbf{61}, e202201083\relax
\mciteBstWouldAddEndPuncttrue
\mciteSetBstMidEndSepPunct{\mcitedefaultmidpunct}
{\mcitedefaultendpunct}{\mcitedefaultseppunct}\relax
\EndOfBibitem
\bibitem[Linden \emph{et~al.}(2011)Linden, Franks, Akbey, Lange, van Rossum,
  and Oschkinat]{Oschkinat2011CryoTResolution}
A.~H. Linden, W.~T. Franks, U.~Akbey, S.~Lange, B.-J. van Rossum and
  H.~Oschkinat, \emph{J. Biomol. NMR}, 2011, \textbf{51}, 283--292\relax
\mciteBstWouldAddEndPuncttrue
\mciteSetBstMidEndSepPunct{\mcitedefaultmidpunct}
{\mcitedefaultendpunct}{\mcitedefaultseppunct}\relax
\EndOfBibitem
\bibitem[Siemer \emph{et~al.}(2012)Siemer, Huang, and
  McDermott]{McDermott2012SSNMRLinewidth}
A.~B. Siemer, K.-Y. Huang and A.~E. McDermott, \emph{PLOS ONE}, 2012,
  \textbf{7}, e47242\relax
\mciteBstWouldAddEndPuncttrue
\mciteSetBstMidEndSepPunct{\mcitedefaultmidpunct}
{\mcitedefaultendpunct}{\mcitedefaultseppunct}\relax
\EndOfBibitem
\bibitem[Jaudzems \emph{et~al.}(2019)Jaudzems, Polenova, Pintacuda, Oschkinat,
  and Lesage]{OschkinatLesage2019DNPbioNMR}
K.~Jaudzems, T.~Polenova, G.~Pintacuda, H.~Oschkinat and A.~Lesage, \emph{J.
  Struct. Biol.}, 2019, \textbf{206}, 90--98\relax
\mciteBstWouldAddEndPuncttrue
\mciteSetBstMidEndSepPunct{\mcitedefaultmidpunct}
{\mcitedefaultendpunct}{\mcitedefaultseppunct}\relax
\EndOfBibitem
\bibitem[Lee \emph{et~al.}(2015)Lee, Hediger, and {De
  Paëpe}]{DePaepe2015NMRenhancedbyDNP}
D.~Lee, S.~Hediger and G.~{De Paëpe}, \emph{Solid State Nucl. Magn. Reson.},
  2015, \textbf{66-67}, 6--20\relax
\mciteBstWouldAddEndPuncttrue
\mciteSetBstMidEndSepPunct{\mcitedefaultmidpunct}
{\mcitedefaultendpunct}{\mcitedefaultseppunct}\relax
\EndOfBibitem
\bibitem[Bordignon \emph{et~al.}(2015)Bordignon, Nalepa, Savitsky, Braun, and
  Jeschke]{bordignon2015_glasses}
E.~Bordignon, A.~I. Nalepa, A.~Savitsky, L.~Braun and G.~Jeschke, \emph{J.
  Phys. Chem. B}, 2015, \textbf{119}, 13797--13806\relax
\mciteBstWouldAddEndPuncttrue
\mciteSetBstMidEndSepPunct{\mcitedefaultmidpunct}
{\mcitedefaultendpunct}{\mcitedefaultseppunct}\relax
\EndOfBibitem
\bibitem[Emmanouilidis \emph{et~al.}(2021)Emmanouilidis, Esteban-Hofer,
  Damberger, de~Vries, Nguyen, Ibanez, Mergenthal, Klotzsch, Yulikov,
  Jeschke,\emph{et~al.}]{emmanouilidis2021nmr}
L.~Emmanouilidis, L.~Esteban-Hofer, F.~F. Damberger, T.~de~Vries, C.~K. Nguyen,
  L.~F. Ibanez, S.~Mergenthal, E.~Klotzsch, M.~Yulikov, G.~Jeschke
  \emph{et~al.}, \emph{Nat. Chem. Biol.}, 2021, \textbf{17}, 608--614\relax
\mciteBstWouldAddEndPuncttrue
\mciteSetBstMidEndSepPunct{\mcitedefaultmidpunct}
{\mcitedefaultendpunct}{\mcitedefaultseppunct}\relax
\EndOfBibitem
\bibitem[Garman and Owen(2006)]{Garman2006}
E.~F. Garman and R.~L. Owen, \emph{Acta Crystallogr. D Biol. Cryst.}, 2006,
  \textbf{62}, 32--47\relax
\mciteBstWouldAddEndPuncttrue
\mciteSetBstMidEndSepPunct{\mcitedefaultmidpunct}
{\mcitedefaultendpunct}{\mcitedefaultseppunct}\relax
\EndOfBibitem
\bibitem[Li \emph{et~al.}(2025)Li, He, Cao, Zhang, Chen, Xiao, and
  Huang]{CryoEM01}
T.~Li, J.~He, H.~Cao, Y.~Zhang, J.~Chen, Y.~Xiao and S.-Y. Huang, \emph{Nat.
  Biotechnol.}, 2025, \textbf{43}, 97--105\relax
\mciteBstWouldAddEndPuncttrue
\mciteSetBstMidEndSepPunct{\mcitedefaultmidpunct}
{\mcitedefaultendpunct}{\mcitedefaultseppunct}\relax
\EndOfBibitem
\bibitem[Liebschner \emph{et~al.}(2019)Liebschner, Afonine, Baker, Bunkoczi,
  Chen, Croll, Hintze, Hung, Jain, McCoy, Moriarty, Oeffner, Poon, Prisant,
  Read, Richardson, Richardson, Sammito, Sobolev, Stockwell, Terwilliger,
  Urzhumtsev, Videau, Williams, and Adams]{CryoEM02}
D.~Liebschner, P.~V. Afonine, M.~L. Baker, G.~Bunkoczi, V.~B. Chen, T.~I.
  Croll, B.~Hintze, L.-W. Hung, S.~Jain, A.~J. McCoy, N.~W. Moriarty, R.~D.
  Oeffner, B.~K. Poon, M.~G. Prisant, R.~J. Read, J.~S. Richardson, D.~C.
  Richardson, M.~D. Sammito, O.~V. Sobolev, D.~H. Stockwell, T.~C. Terwilliger,
  A.~G. Urzhumtsev, L.~L. Videau, C.~J. Williams and P.~D. Adams, \emph{Acta
  Crystallogr. D Struct. Biol.}, 2019, \textbf{75}, 861--877\relax
\mciteBstWouldAddEndPuncttrue
\mciteSetBstMidEndSepPunct{\mcitedefaultmidpunct}
{\mcitedefaultendpunct}{\mcitedefaultseppunct}\relax
\EndOfBibitem
\bibitem[Robinson~Brown \emph{et~al.}(2024)Robinson~Brown, Webber, Casey,
  Franck, Shell, and Han]{SongiHan2024ODNP_SolventRestructuring}
D.~C. Robinson~Brown, T.~R. Webber, T.~M. Casey, J.~Franck, M.~S. Shell and
  S.~Han, \emph{Phys. Chem. Chem. Phys.}, 2024, \textbf{26}, 14637 –
  14650\relax
\mciteBstWouldAddEndPuncttrue
\mciteSetBstMidEndSepPunct{\mcitedefaultmidpunct}
{\mcitedefaultendpunct}{\mcitedefaultseppunct}\relax
\EndOfBibitem
\bibitem[Moon \emph{et~al.}(2024)Moon, Webber, Brown, Richardson, Casey,
  Segalman, Shell, and Han]{SongiHan2024WaterPolymerDiffusivity}
J.~D. Moon, T.~R. Webber, D.~R. Brown, P.~M. Richardson, T.~M. Casey, R.~A.
  Segalman, M.~S. Shell and S.~Han, \emph{Chem. Sci.}, 2024, \textbf{15}, 2495
  – 2508\relax
\mciteBstWouldAddEndPuncttrue
\mciteSetBstMidEndSepPunct{\mcitedefaultmidpunct}
{\mcitedefaultendpunct}{\mcitedefaultseppunct}\relax
\EndOfBibitem
\bibitem[Das~Mahanta \emph{et~al.}(2023)Das~Mahanta, Robinson~Brown, Pezzotti,
  Han, Schwaab, Shell, and Havenith]{Havenith2023LocalSolvationWaterGlycerol}
D.~Das~Mahanta, D.~Robinson~Brown, S.~Pezzotti, S.~Han, G.~Schwaab, M.~S. Shell
  and M.~Havenith, \emph{Chem. Sci.}, 2023, \textbf{14}, 7381 – 7392\relax
\mciteBstWouldAddEndPuncttrue
\mciteSetBstMidEndSepPunct{\mcitedefaultmidpunct}
{\mcitedefaultendpunct}{\mcitedefaultseppunct}\relax
\EndOfBibitem
\bibitem[Bag \emph{et~al.}(2024)Bag, Dec, Pezzotti, Sahoo, Schwaab, Winter, and
  Havenith]{Havenith2024HydrationFibrils}
S.~Bag, R.~Dec, S.~Pezzotti, R.~R. Sahoo, G.~Schwaab, R.~Winter and
  M.~Havenith, \emph{Biophys. J.}, 2024, \textbf{123}, 3863 – 3870\relax
\mciteBstWouldAddEndPuncttrue
\mciteSetBstMidEndSepPunct{\mcitedefaultmidpunct}
{\mcitedefaultendpunct}{\mcitedefaultseppunct}\relax
\EndOfBibitem
\bibitem[König \emph{et~al.}(2024)König, Pezzotti, Ramos, Schwaab, and
  Havenith]{Havenith2024SolvationLLPS}
B.~König, S.~Pezzotti, S.~Ramos, G.~Schwaab and M.~Havenith, \emph{Biophys.
  J.}, 2024, \textbf{123}, 1367 – 1375\relax
\mciteBstWouldAddEndPuncttrue
\mciteSetBstMidEndSepPunct{\mcitedefaultmidpunct}
{\mcitedefaultendpunct}{\mcitedefaultseppunct}\relax
\EndOfBibitem
\bibitem[König \emph{et~al.}(2025)König, Pezzotti, Schwaab, and
  Havenith]{Havenith2025TuningByCoSolutes}
B.~König, S.~Pezzotti, G.~Schwaab and M.~Havenith, \emph{Chem. Sci.}, 2025,
  \textbf{16}, 5897 – 5906\relax
\mciteBstWouldAddEndPuncttrue
\mciteSetBstMidEndSepPunct{\mcitedefaultmidpunct}
{\mcitedefaultendpunct}{\mcitedefaultseppunct}\relax
\EndOfBibitem
\bibitem[Angarita \emph{et~al.}(2021)Angarita, Mazzobre, Corti, and
  Longinotti]{angarita2021revisiting}
I.~Angarita, M.~F. Mazzobre, H.~R. Corti and M.~P. Longinotti, \emph{Phys.
  Chem. Chem. Phys.}, 2021, \textbf{23}, 17018--17025\relax
\mciteBstWouldAddEndPuncttrue
\mciteSetBstMidEndSepPunct{\mcitedefaultmidpunct}
{\mcitedefaultendpunct}{\mcitedefaultseppunct}\relax
\EndOfBibitem
\bibitem[B{\"o}hmer \emph{et~al.}(2024)B{\"o}hmer, Gabriel, Costigliola,
  Kociok, Hecksher, Dyre, and Blochowicz]{bohmer2024time}
T.~B{\"o}hmer, J.~P. Gabriel, L.~Costigliola, J.-N. Kociok, T.~Hecksher, J.~C.
  Dyre and T.~Blochowicz, \emph{Nat. Phys.}, 2024,  1--9\relax
\mciteBstWouldAddEndPuncttrue
\mciteSetBstMidEndSepPunct{\mcitedefaultmidpunct}
{\mcitedefaultendpunct}{\mcitedefaultseppunct}\relax
\EndOfBibitem
\bibitem[Amir \emph{et~al.}(2012)Amir, Oreg, and Imry]{amir2012relaxations}
A.~Amir, Y.~Oreg and Y.~Imry, \emph{Proc. Natl. Acad. Sci.}, 2012,
  \textbf{109}, 1850--1855\relax
\mciteBstWouldAddEndPuncttrue
\mciteSetBstMidEndSepPunct{\mcitedefaultmidpunct}
{\mcitedefaultendpunct}{\mcitedefaultseppunct}\relax
\EndOfBibitem
\bibitem[Murata and Tanaka(2012)]{murata2012liquid}
K.-i. Murata and H.~Tanaka, \emph{Nat. Mater.}, 2012, \textbf{11},
  436--443\relax
\mciteBstWouldAddEndPuncttrue
\mciteSetBstMidEndSepPunct{\mcitedefaultmidpunct}
{\mcitedefaultendpunct}{\mcitedefaultseppunct}\relax
\EndOfBibitem
\bibitem[Kuzin \emph{et~al.}(2022)Kuzin, Jeschke, and
  Yulikov]{kuzin2022diffusion}
S.~Kuzin, G.~Jeschke and M.~Yulikov, \emph{Phys. Chem. Chem. Phys.}, 2022,
  \textbf{24}, 23517--23531\relax
\mciteBstWouldAddEndPuncttrue
\mciteSetBstMidEndSepPunct{\mcitedefaultmidpunct}
{\mcitedefaultendpunct}{\mcitedefaultseppunct}\relax
\EndOfBibitem
\bibitem[Kuzin \emph{et~al.}(2024)Kuzin, Stolba, Wu, Syryamina, Boulos,
  Jeschke, Nystr\"om, and Yulikov]{kuzin2024quantification}
S.~Kuzin, D.~Stolba, X.~Wu, V.~N. Syryamina, S.~Boulos, G.~Jeschke,
  L.~Nystr\"om and M.~Yulikov, \emph{J. Phys. Chem. Lett.}, 2024, \textbf{15},
  5625--5632\relax
\mciteBstWouldAddEndPuncttrue
\mciteSetBstMidEndSepPunct{\mcitedefaultmidpunct}
{\mcitedefaultendpunct}{\mcitedefaultseppunct}\relax
\EndOfBibitem
\bibitem[Kuzin \emph{et~al.}(2025)Kuzin, Syryamina, Qi, Fischer, H{\"u}lsmann,
  Godt, Jeschke, and Yulikov]{kuzin2025ihridme}
S.~Kuzin, V.~N. Syryamina, M.~Qi, M.~Fischer, M.~H{\"u}lsmann, A.~Godt,
  G.~Jeschke and M.~Yulikov, \emph{Magn. Reson.}, 2025, \textbf{6},
  93--112\relax
\mciteBstWouldAddEndPuncttrue
\mciteSetBstMidEndSepPunct{\mcitedefaultmidpunct}
{\mcitedefaultendpunct}{\mcitedefaultseppunct}\relax
\EndOfBibitem
\bibitem[Kuzin and Yulikov(2025)]{kuzin2025ridme}
S.~Kuzin and M.~Yulikov, \emph{J. Phys. Chem. Lett.}, 2025, \textbf{16},
  1024--1037\relax
\mciteBstWouldAddEndPuncttrue
\mciteSetBstMidEndSepPunct{\mcitedefaultmidpunct}
{\mcitedefaultendpunct}{\mcitedefaultseppunct}\relax
\EndOfBibitem
\bibitem[Milikisyants \emph{et~al.}(2009)Milikisyants, Scarpelli, Finiguerra,
  Ubbink, and Huber]{milikisyants2009pulsed}
S.~Milikisyants, F.~Scarpelli, M.~G. Finiguerra, M.~Ubbink and M.~Huber,
  \emph{J. Magn. Reson.}, 2009, \textbf{201}, 48--56\relax
\mciteBstWouldAddEndPuncttrue
\mciteSetBstMidEndSepPunct{\mcitedefaultmidpunct}
{\mcitedefaultendpunct}{\mcitedefaultseppunct}\relax
\EndOfBibitem
\bibitem[Keller \emph{et~al.}(2019)Keller, Qi, Gmeiner, Ritsch, Godt, Jeschke,
  Savitsky, and Yulikov]{keller2019intermolecular}
K.~Keller, M.~Qi, C.~Gmeiner, I.~Ritsch, A.~Godt, G.~Jeschke, A.~Savitsky and
  M.~Yulikov, \emph{Phys. Chem. Chem. Phys.}, 2019, \textbf{21},
  8228--8245\relax
\mciteBstWouldAddEndPuncttrue
\mciteSetBstMidEndSepPunct{\mcitedefaultmidpunct}
{\mcitedefaultendpunct}{\mcitedefaultseppunct}\relax
\EndOfBibitem
\bibitem[de~Sousa and Das~Sarma(2003)]{deSousa2003}
R.~de~Sousa and S.~Das~Sarma, \emph{Phys. Rev. B}, 2003, \textbf{68},
  115322\relax
\mciteBstWouldAddEndPuncttrue
\mciteSetBstMidEndSepPunct{\mcitedefaultmidpunct}
{\mcitedefaultendpunct}{\mcitedefaultseppunct}\relax
\EndOfBibitem
\bibitem[Witzel and Das~Sarma(2006)]{Witzel2006}
W.~M. Witzel and S.~Das~Sarma, \emph{Phys. Rev. B}, 2006, \textbf{74},
  035322\relax
\mciteBstWouldAddEndPuncttrue
\mciteSetBstMidEndSepPunct{\mcitedefaultmidpunct}
{\mcitedefaultendpunct}{\mcitedefaultseppunct}\relax
\EndOfBibitem
\bibitem[Jahn \emph{et~al.}(2022)Jahn, Canarie, and Stoll]{jahn2022mechanism}
S.~M. Jahn, E.~R. Canarie and S.~Stoll, \emph{J. Phys. Chem. Lett.}, 2022,
  \textbf{13}, 5474--5479\relax
\mciteBstWouldAddEndPuncttrue
\mciteSetBstMidEndSepPunct{\mcitedefaultmidpunct}
{\mcitedefaultendpunct}{\mcitedefaultseppunct}\relax
\EndOfBibitem
\bibitem[Jeschke(2023)]{jeschke2023nuclear}
G.~Jeschke, \emph{J. Magn. Reson. Open}, 2023, \textbf{14}, 100094\relax
\mciteBstWouldAddEndPuncttrue
\mciteSetBstMidEndSepPunct{\mcitedefaultmidpunct}
{\mcitedefaultendpunct}{\mcitedefaultseppunct}\relax
\EndOfBibitem
\bibitem[Kuzin \emph{et~al.}(2024)Kuzin, Yulikov, and
  Jeschke]{kuzin2024perturbation}
S.~Kuzin, M.~Yulikov and G.~Jeschke, \emph{J. Magn. Reson.}, 2024,
  \textbf{365}, 107729\relax
\mciteBstWouldAddEndPuncttrue
\mciteSetBstMidEndSepPunct{\mcitedefaultmidpunct}
{\mcitedefaultendpunct}{\mcitedefaultseppunct}\relax
\EndOfBibitem
\bibitem[Tschaggelar \emph{et~al.}(2009)Tschaggelar, Kasumaj, Santangelo,
  Forrer, Leger, Dube, Diederich, Harmer, Schuhmann, García-Rubio, and
  Jeschke]{TSCHAGGELAR200981}
R.~Tschaggelar, B.~Kasumaj, M.~G. Santangelo, J.~Forrer, P.~Leger, H.~Dube,
  F.~Diederich, J.~Harmer, R.~Schuhmann, I.~García-Rubio and G.~Jeschke,
  \emph{J. Magn. Reson.}, 2009, \textbf{200}, 81--87\relax
\mciteBstWouldAddEndPuncttrue
\mciteSetBstMidEndSepPunct{\mcitedefaultmidpunct}
{\mcitedefaultendpunct}{\mcitedefaultseppunct}\relax
\EndOfBibitem
\bibitem[Keller \emph{et~al.}(2016)Keller, Doll, Qi, Godt, Jeschke, and
  Yulikov]{keller2016averaging}
K.~Keller, A.~Doll, M.~Qi, A.~Godt, G.~Jeschke and M.~Yulikov, \emph{J. Magn.
  Reson.}, 2016, \textbf{272}, 108--113\relax
\mciteBstWouldAddEndPuncttrue
\mciteSetBstMidEndSepPunct{\mcitedefaultmidpunct}
{\mcitedefaultendpunct}{\mcitedefaultseppunct}\relax
\EndOfBibitem
\bibitem[Zecevic \emph{et~al.}(1998)Zecevic, Eaton, Eaton, and
  Lindgren]{Eatons1998EPRDephasingbyProtons}
A.~Zecevic, G.~Eaton, S.~Eaton and M.~Lindgren, \emph{Mol. Phys.}, 1998,
  \textbf{95}, 1255--1263\relax
\mciteBstWouldAddEndPuncttrue
\mciteSetBstMidEndSepPunct{\mcitedefaultmidpunct}
{\mcitedefaultendpunct}{\mcitedefaultseppunct}\relax
\EndOfBibitem
\bibitem[El~Mkami \emph{et~al.}(2014)El~Mkami, Ward, Bowman, Owen-Hughes, and
  Norman]{ElMkami2014ProtDeut}
H.~El~Mkami, R.~Ward, A.~Bowman, T.~Owen-Hughes and D.~G. Norman, \emph{J.
  Magn. Reson.}, 2014, \textbf{248}, 36--41\relax
\mciteBstWouldAddEndPuncttrue
\mciteSetBstMidEndSepPunct{\mcitedefaultmidpunct}
{\mcitedefaultendpunct}{\mcitedefaultseppunct}\relax
\EndOfBibitem
\bibitem[Ward \emph{et~al.}(2010)Ward, Bowman, Sozudogru, El-Mkami,
  Owen-Hughes, and Norman]{Ward2010DistanceMeasurementsDeutProt}
R.~Ward, A.~Bowman, E.~Sozudogru, H.~El-Mkami, T.~Owen-Hughes and D.~G. Norman,
  \emph{J. Magn. Reson.}, 2010, \textbf{207}, 164--167\relax
\mciteBstWouldAddEndPuncttrue
\mciteSetBstMidEndSepPunct{\mcitedefaultmidpunct}
{\mcitedefaultendpunct}{\mcitedefaultseppunct}\relax
\EndOfBibitem
\bibitem[gly()]{glycerolGeometry}
\url{https://pubchem.ncbi.nlm.nih.gov/compound/Glycerol}, Accessed July 26,
  2025\relax
\mciteBstWouldAddEndPuncttrue
\mciteSetBstMidEndSepPunct{\mcitedefaultmidpunct}
{\mcitedefaultendpunct}{\mcitedefaultseppunct}\relax
\EndOfBibitem
\bibitem[Weinberg and Zimmerman(1955)]{weinberg1955}
I.~Weinberg and J.~R. Zimmerman, \emph{J. Chem. Phys.}, 1955, \textbf{23},
  748--749\relax
\mciteBstWouldAddEndPuncttrue
\mciteSetBstMidEndSepPunct{\mcitedefaultmidpunct}
{\mcitedefaultendpunct}{\mcitedefaultseppunct}\relax
\EndOfBibitem
\bibitem[Eggeling \emph{et~al.}(2023)Eggeling, Soetbeer,
  F{\'a}bregas-Ib{\'a}{\~n}ez, Klose, and Jeschke]{eggeling2023quantifying}
A.~Eggeling, J.~Soetbeer, L.~F{\'a}bregas-Ib{\'a}{\~n}ez, D.~Klose and
  G.~Jeschke, \emph{Phys. Chem. Chem. Phys.}, 2023, \textbf{25},
  11145--11157\relax
\mciteBstWouldAddEndPuncttrue
\mciteSetBstMidEndSepPunct{\mcitedefaultmidpunct}
{\mcitedefaultendpunct}{\mcitedefaultseppunct}\relax
\EndOfBibitem
\bibitem[Popov \emph{et~al.}(2015)Popov, Greenbaum~(Gutina), Sokolov, and
  Feldman]{Feldman2015wgice}
I.~Popov, A.~Greenbaum~(Gutina), A.~P. Sokolov and Y.~Feldman, \emph{Phys.
  Chem. Chem. Phys.}, 2015, \textbf{17}, 18063--18071\relax
\mciteBstWouldAddEndPuncttrue
\mciteSetBstMidEndSepPunct{\mcitedefaultmidpunct}
{\mcitedefaultendpunct}{\mcitedefaultseppunct}\relax
\EndOfBibitem
\end{mcitethebibliography}
\bibliographystyle{rsc} %the RSC's .bst file

\end{document}